\documentclass{aastex}
\usepackage{graphicx}
\begin{document}
\def\lam{$\lambda$}
\def\kms{km~s$^{-1}$}
\def\vphot{$v_{phot}$}
\def\ang{~\AA}

\title{Optical Spectra of the Type~Ia Supernova 1998aq}

\author{David Branch\altaffilmark{1}, Peter Garnavich\altaffilmark{2},
Thomas~Matheson\altaffilmark{3}, E.~Baron\altaffilmark{1},
R.~C.~Thomas\altaffilmark{1}, Kazuhito~Hatano\altaffilmark{4},
Peter~Challis\altaffilmark{3}, S.~Jha\altaffilmark{3}, and
Robert~P.~Kirshner\altaffilmark{3}}

\altaffiltext{1}{Dept. of Physics and Astronomy, University of Oklahoma,
Norman, OK 73019}

\altaffiltext{2}{Dept. of Physics and Astronomy, University of Notre Dame,
Notre Dame, IN 46556}

\altaffiltext{3}{Harvard--Smithsonian Center for Astrophysics, 60 Garden
Street, Cambridge, MA 02138}

\altaffiltext{4}{Department of Astronomy, School of Science, University
of Tokyo, Tokyo, Japan}

\begin{abstract} 

We present 29 optical spectra of the normal Type~Ia SN~1998aq, ranging
from 9~days before to 241~days after the time of maximum brightness.
This spectrocopic data set, together with photometric data presented
elsewhere, makes SN~1998aq one of the best observed Type~Ia supernovae
at optical wavelengths.  We use the parameterized supernova
synthetic--spectrum code {\bf Synow} to study line identifications in
the early photospheric--phase spectra.  The results include evidence
for lines of singly ionized carbon, at ejection velocities as low as
11,000 \kms.  Implications for SN~Ia explosion models are discussed.

\end{abstract}


\section{INTRODUCTION}

The discovery in a relatively nearby galaxy of a Type~Ia supernova
(SN~Ia), more than a week before its time of maximum brightness,
presents an observational opportunity that should not be missed.  A
few days after the discovery of SN 1998aq in NGC 3982 on 1998 April 13
by the U.K. Supernova/Nova Patrol (Hurst, Armstrong, \& Arbour 1998),
photometric and spectroscopic observations undertaken at the
F.~L.~Whipple Observatory (FLWO) revealed that the supernova was a
Type~Ia whose brightness was on the rise (Garnavich et~al. 1998).
Subsequent observations resulted in SN~1998aq becoming one of the best
observed SNe~Ia at optical wavelengths.  The {\sl Hubble Space
Telescope} has been used by Saha et~al (2001; see also Stetson
\& Gibson 2001) to make a Cepheid--based determination of the distance
to NGC~3982, so that SN~1998aq can be added to the list of SNe~Ia that
are used to calibrate the extragalactic distance scale.

A preliminary report of the photometric observations of SN~1998aq has
been given by Boffi \& Riess (2003), and the final results will appear
elsewhere (A.~G.~Riess et~al., in preparation).  SN~1998aq reached
a maximum brightness of $B$=12.39 on 1998 April~27.  The value of
$\Delta m_{15}$, the decline of the blue--band magnitude during the
first 15 days after maximum (Phillips 1993), was 1.14, a typical value
for a normal SN~Ia. SN~1998aq was photometrically normal, except that
some of its broad--band colors were unusually blue, e.g., $B-V$=$-$0.18
at the time of $B$ maximum.

\section{OBSERVATIONS}

Low--dispersion spectra of SN~1998aq were obtained with the FAST
spectrograph (Fabricant et~al. 1998) on the 1.5-m Tillinghast
telescope at FLWO.  The FAST spectrograph uses a 2688$\times$512 Loral
CCD with a spatial scale of 1.$\arcsec$1 per pixel in the binning mode
used for these observations.  Details of the exposures are given in
Table~1.  The data were reduced in the standard manner with
IRAF\footnote{IRAF is distributed by the National Optical Astronomy
Observatories, which are operated by the Association of Universities
for Research in Astronomy, Inc., under cooperative agreement with the
National Science Foundation.} and our own routines.  The spectra were
optimally extracted (Horne 1986).  Wavelength calibration was
accomplished with HeNeAr lamps taken immediately after each SN
exposure.  Small--scale adjustments derived from night--sky lines in the
SN frames were also applied.  Spectrophotometric standards are listed
in Table~1.  We attempted to remove telluric lines using the
well--exposed continua of the spectrophotometric standards (Wade \&
Horne 1988; Matheson et~al. 2001).  The spectra were, in general, not
taken at the parallactic angle (Filippenko 1982).  For the spectra
modeled in this paper, though, the airmass was low ($< 1.2$).

The spectra, labelled by the epoch in days with respect to the date of
maximum brightness in the $B$ band (1998 April~27), are displayed in
Figs. 1 - 3.  The spectra are those of a normal SN~Ia.  The early
spectra, from $-9$ to 7 days (Fig.~1), contain the characteristic deep
Si~II absorption near 6100~\AA\ and the distinctive S~II W--shaped
feature from about 5200 to 5500~\AA.  The spectra between 19 and 91
days (Figs.~2 and 3) are those of a SN~Ia making the transformation
from the photospheric to the nebular phase.  By 211 days (Fig.~3) the
transformation was complete.

In Figs.~4 and 5, four spectra of SN~1998aq are compared to those of
another well observed event, SN~1994D (Patat et al. 1996).  The
similarity of the spectra at $-8$ days (Fig.~4) is impressive,
considering that pre--maximum spectra of SNe~Ia exhibit considerable
diversity (e.g., Hatano et~al. 2000; Li et~al. 2001).  At $-8$ days
the main difference is that SN~1998aq has a peak near 4600~\AA\ that
is not present in SN~1994D; otherwise practically all of the features,
including the weak ones, are quite similar.  At 4, 19, and 54/55 days
the spectra also are similar, although they are distinguishable, e.g.,
near 4100~\AA\ at 4 days and near 5400~\AA\ at 19 and 54/55 days.

In Fig.~6 the blueshifts, expressed in \kms, of the absorption near
6100~\AA\ produced by Si~II \lam6355 in the early spectra of
SNe~1998aq and 1994D are compared. Except at the earliest times, when
the blueshift is changing rapidly with time and that of SN~1994D is
mildly higher than that of SN~1998aq, the differences are within the
measurement errors of about 250 \kms.  Fig.~7 shows the blueshifts of
four absorption features in SN~1998aq that are not badly blended, and
of which we are confident of the identifications (Si~II \lam6355,
Ca~II \lam3945, S~II \lam5654, and Si~III \lam4560).  The blueshifts
of the deep Ca~II and Si~II absorptions remain near 10,000 \kms\ even
at 35 days, but those of the weaker Si~III and S~II features decrease
more rapidly.

From Fig.~6 we estimate a value of 10,300 \kms\ for the
$v_{10}(Si~II)$ parameter --- the blueshift of the deep Si~II
absorption 10 days after maximum brightness (Branch \& van den Bergh
1993).  In the 0, 1, 2, and 3~day spectra we measure values of 0.22,
0.20, 0.22, and 0.24 for the parameter $R(Si~II)$ (Nugent
et~al. 1995), in good agreement with the values of $0.22 \pm 0.02$
measured by Vinko et~al. (1999) in a $-5$~day spectrum; we recommend
$R(Si~II)$=0.22$\pm$0.01.  Hatano et~al. (2000) showed and discussed
the diversity among SNe~Ia in a plot of $R(Si~II)$ versus
$v_{10}(Si~II)$ (their Fig.~1).  Our measured values put SN~1998aq in
the most heavily populated part of their plot, near, e.g., SNe~1994D,
1996X, 1998bu, and 1990N.  Garnavich et~al. (2000) have shown that
there is a tight relation between $R(Si~II)$ [referred to as
$R(Si~II/Ti~II)$ by Garnavich et~al.] and $\Delta m_{15}$.  Garnavich
et~al. intended to measure $R(Si~II)$ only in spectra obtained within
three days of maximum light, but for SN~1998aq their adopted date of
maximum differed from the date later reported by Boffi \& Riess
(2003). With our recommended value of $R(Si~II)$, based on specta
obtained within three days of the revised date of maximum, SN~1998aq
fits the Garnavich et~al. relation between $R(Si~II)$ and $\Delta
m_{15}$ even better.

\section{LINE IDENTIFICATIONS IN EARLY SPECTRA}

The analysis of supernova spectra begins with the often difficult
process of making line identifications.  Only a limited number of good
SN~Ia spectra have been obtained well before the time of maximum
light, so we are especially interested in exploring line
identifications in the earliest spectrum of SN~1998aq, at $-9$ days.
For comparison we also will look at the 0 and 7~day spectra.  These
three spectra are compared in Fig.~8.  The overall appearance of these
spectra is rather similar, although some significant changes do occur
during this time interval: the Si~II, S~II, and Ca~II absorptions
become deeper with time; the absorption trough from 4800 to 5000~\AA\
in the $-9$~day spectrum breaks up in the 7~day spectrum; and the
absorption trough from 4200 to 4500~\AA\ in the 7~day spectrum is more
broken up in the $-9$~day spectrum.

We have studied line identifications by comparing the three spectra of
Fig.~8 with numerous synthetic spectra generated with the
parameterized supernova synthetic--spectrum code {\bf Synow}.  This
code has been used and described in several recent papers, e.g.,
Branch et~al. (2002).  The basic assumptions are spherical symmetry, a
sharp photosphere, velocity proportional to radius, and resonance
scattering in the Sobolev approximation.  The shape of the underlying
continuum is that of a blackbody at temperature $T_{bb}$.  Line
optical depths are taken to vary as $\exp(-v/v_e)$, with $v_e$=1000
\kms\ for all synthetic spectra shown in this paper.  For each ion
whose lines are introduced, the maximum Sobolev optical depth of a
reference line is a fitting parameter.  Ordinarily, the maximum line
optical depth is at the photosphere, but for a ``detached'' ion the
line optical depth is zero at the photosphere and rises
discontinuously to its maximum value at some higher detachment
velocity.  The optical depths of the other lines of an ion, relative
to that of the reference line, are calculated for Boltzmann excitation
at temperature $T_{exc}$. In this paper, instead of adopting a common
value of the excitation temperature $T_{exc}$ as the default value,
the default value for each ion is the temperature at which the ion's
reference--line optical depth reaches a maximum in plots of optical
depth versus temperature (Hatano et~al. 1999b).  This is a reasonable
default since, e.g., if lines of high excitation are present they are
likely to be formed in high temperature layers.  The default
excitation temperatures are used throughout this paper except in one
instance that is mentioned below.

\subsection{The $-9$ Day Spectrum}

The $-9$ day spectrum is shown in Fig.~9, with the measured
wavelengths of absorption features labelled, so that we can
conveniently refer to each absorption throughout this subsection. In
this and all subsequent figures of this paper, instead of plotting
$f_{\nu}$ on the vertical axis we plot $f_{\nu}/{\nu}$, to make the
spectrum approximately flat.  This provides a better view of features
at the longer wavelengths where the underlying continuum is low, which
is helpful because the diagnostic value of a spectral feature is not
proportional to the level of the underlying continuum.  Note also that
in order to show as much detail as possible, in this and subsequent
figures the flux axis does not go to zero.

In Fig.~10 the same $-9$ day spectrum is compared with a synthetic
spectrum that has $v_{phot}=13,000$ \kms, $T_{bb}$=16,000~K, and
includes lines of seven ions.  The ion--specific parameters of the
synthetic spectrum are listed in Table~2.  The fit shown in Fig.~10 is
good by supernova standards, although there are two significant
discrepancies that we have not yet resolved: the synthetic spectrum is
too high from 4460 to 4580~\AA, and too low from 4000 to 4160~\AA.
The contributions of each ion to the synthetic spectrum of Fig.~10 are
shown in Figs.~11 to 17, each of which we now briefly discuss.

Fig.~11 shows that the main contribution of Si~II is that
$\lambda$6355 accounts for almost all of the 6115~\AA\ absorption.
Note that $\lambda$6355 produces a synthetic emission peak where an
absorption feature is observed.  The \lam5972 and \lam5051 lines
contribute to the 5740 and 4870~\AA\ absorptions.  The \lam4130 line
contributes to the 3970~\AA\ absorption, but absorption by \lam4130
also is partially responsible for our failure to fit the observed
emission peak near 4070\ang.  In any case, the presence of Si~II
\lam6355 in the $-9$ day spectrum of SN~1998aq is, of course,
definite.

Fig.~12 shows that the main contribution of S~II is that \lam5654 and
5468 produce the 5420 and 5260\ang\ absorptions. Lines near \lam6305
make a minor contribution to the 6115\ang\ absorption.  The \lam5208
line may be responsible for the 5110\ang\ absorption, although it does
not produce a distinct absorption in the full synthetic spectrum of
Fig.~10.  The \lam5032 line contributes to the 4870\ang\ absorption
and some weaker S~II lines make minor contributions.  The \lam4163
line is partially responsible for our failure to fit the observed
emission peak near 4070\ang.  Nevertheless, the presence of S~II
\lam5654 and 5468 is definite.

Fig.~13 shows that the only contribution of Ca~II (since the spectrum
does not extend to the infrared triplet) is that \lam3945 (the H\&K
lines) produces the 3805\ang\ absorption.  The presence of Ca~II
\lam3945 is definite.

Fig.~14 shows that Si~III \lam4560 produces the 4390\ang\ absorption.
It is likely that \lam5743 is responsible for the 5515\ang\
absorption, although in Fig.~10 the synthetic absorption is at a
slightly shorter wavelength that the observed one.  We are confident
of the presence of Si~III \lam4560.

Fig.~15 shows that Fe~III \lam5156 and 4420 are largely responsible
for the 4930 and 4255\ang\ absorptions.  Lines near \lam6000 also
contribute to the 5740\ang\ absorption.  This is the one instance in
this paper where we depart from the default excitation temperature.
At the Fe~III default value of 14,000~K the feature produced by
\lam6000 is present but weaker.  We use 16,000~K to show that Fe~III
might be mainly responsible for the 5740\ang\ absorption at this
epoch.  In any case, we are confident of the presence of Fe~III
\lam5156 and 4420.

Fig.~16 shows the contribution of Fe~II lines, detached at 20,000
\kms.  This means that in the synthetic spectrum, the Fe~II optical
depths are zero between the velocity at the photosphere, 13,000 \kms,
and the detachment velocity of 20,000 \kms.  The presence of detached,
high--velocity Fe~II in SN~1994D was suggested by Hatano et
al. (1999a) on the basis of {\bf Synow} synthetic spectra and then
supported by Lentz et~al. (2001b) who carried out
non--local--thermodynamic--equilibrium (NLTE) calculations with the
\texttt{PHOENIX} code.  In Fig.~16, we see that Fe~II \lam5169, 5018, and
4924, are contributing to the 4870, 4730, and 4640\ang\ absorptions,
respectively.  Lines near \lam4549 and 4025 contribute to the 4295 and
3970\ang\ absorptions.  High--velocity Fe~II makes a positive
contribution to the full fit in Fig.~10, and we consider its presence
in the observed spectrum to be likely.

Fig.~17 shows the contributions of C~II lines, slightly detached at
14,000 \kms. The strongest line, \lam6580, is playing an important
role in the synthetic spectrum: it beats down the unwanted emission
component of Si~II \lam6355, and combines with it in the full
synthetic spectrum of Fig.~10 to produce a feature near the observed
6310\ang\ absorption.  The \lam7234 line produces the 6940\ang\
absorption.  (Fig.~17 is a good example of the benefit of plotting
$f_\nu/\nu$: it gives a better view of the features produced by
\lam6580 and \lam7234.)  The \lam4267 line is at the right wavelength
to be consistent with the 4120\ang\ absorption, although \lam4267 does
not produce a distinct feature in the full synthetic spectrum of
Fig.~10.  We consider the presence of C~II lines to be likely.

\subsection{The 0 and 7~Day Spectra}

In Fig.~18 the 0~day spectrum of SN~1998aq is compared with a
synthetic spectrum that has \vphot=11,000 \kms, $T_{bb}$=16,000~K, and
contains lines of the same seven ions that were used for the $-9$~day
spectrum. The ion--specific parameters of the synthetic spectrum are
listed in Table~3.  The Fe~II lines now are detached at 18,000 rather
than 20,000 \kms, and Si~II and S~II are mildly detached at 12,000
\kms.  Compared to the synthetic spectrum for $-9$~days, the Ca~II
line optical depths are increased by a factor of five, those of Si~II
and Fe~III are up by small factors, the Fe~II optical depths are
unchanged, and those of S~II, Si~III, and C~II are down by small
factors.  The overall fit in Fig.~18 is good, the main discrepancies
being that the synthetic spectrum again lacks the 4070\ang\ emission
peak, and now the synthetic spectrum is too low from 5800 to 6000\ang.
The difference in the heights of the apparent continuum levels at
wavelengths longer than 6500\ang\ has been encountered in previous
{\bf Synow} studies of SNe~Ia and probably is caused by our use of a
blackbody continuum.

In Fig.~18, C~II \lam6580 again beats down the Si~II \lam6355 emission
and produces an absorption notch that may correspond to an observed
feature.  The \lam7234 line produces a wiggle in the synthetic
spectrum that may be at least partially responsible for the knee in
the observed spectrum near 7000\ang.

In Fig.~19 the 7~day spectrum of SN~1998aq is compared with a
synthetic spectrum that has \vphot=11,000 \kms\ and $T_{bb}$=16,000~K.
The ion--specific parameters of the synthetic spectrum are listed in
Table~4.  This synthetic spectrum contains lines of six of the seven
ions used previously, Si~III now being omitted.  Fe~II has two
components: one detached at 18,000 \kms\ as in the 0~day spectrum, and
now also an undetached component.  Compared to the 0~day spectrum, the
optical depths of the Si~II, S~II, Ca~II, C~II, and Fe~II lines have
been increased while Fe~III has been decreased.  The synthetic
spectrum of Fig.~19 also includes contributions from Mg~II, Na~I,
Co~II, and Ti~II lines, but their contributions are not major and we
do not consider their presence in the observed spectrum to be
established. To the extent to which we have been able to account for
it, the breakup of the 4800 to 5000\ang\ region and the filling in of
the 4200 to 4500\ang\ region, both mentioned above in the discussion
of Fig.~8, are caused by the emergence of the undetached Fe~II
component.  The main discrepancy in Fig.~19 is that the synthetic
spectrum is too low in the interval 5100 to 6000\ang.

In Fig.~19, C~II \lam6580 leads to a (perhaps fortuitously) good fit
at the top of the Si~II \lam 6355 emission, and again \lam 7234
produces a feature that appears to be responsible for the observed one
near 7000\ang.

\section{DISCUSSION}

The identification of lines of Si~II, S~II, Ca~II, Fe~III, and Si~III
in early spectra of SNe~Ia is not new, but the figures of this paper
clearly illustrate the extent to which lines of these ions can
account, with simple assumptions, for the observed spectral features
in SN~1998aq.  It is instructive to compare parameterized {\bf Synow}
spectrum fits with NLTE synthetic spectra for specific explosion
models.  Lentz et~al. (2001b) used the \texttt{PHOENIX} code to make
detailed calculations of spectra of the deflagration model W7 (Nomoto,
Thielemann, \& Yokoi 1984) and found reasonably good agreement with
the early spectra of SN~1994D, which as we have seen in \S2 was
spectroscopically similar to SN~1998aq.  Fig.~20 is like Fig.~10 but
also including the \texttt{PHOENIX} spectrum of model~W7 at 11 days
after explosion (corresponding to about 9 days before maximum light).
The model~W7 spectrum does have peaks and dips at some of the right
wavelengths, but there is room for improvement.  For example, the W7
spectrum lacks the 4390~\ang\ absorption that we attribute to Si~III
\lam4560, and the W7 emission peaks near 5900 and 6400\ang\ are too
high.  Ongoing work involving close comparative studies of observed
spectra, {\bf Synow}--level synthetic spectra, and
\texttt{PHOENIX}--level spectra of explosion models will provide
indications of how explosion models need to be altered to give better
agreement with observed spectra.

Apparently, Fe~II features forming at high velocity, $\ge 18,000$
\kms, are present at the three epochs we have studied.  Hatano
et~al. (1999a) identified high--velocity ($\ge 20,000$ \kms) Ca~II and
Fe~II lines in SN~1994D.  Kasen et~al. (2003) discuss high--velocity
(18,000 to 25,000 \kms) components of the Ca~II infrared triplet in
flux and polarization spectra of SNe~2001el, and Thomas et~al. (2003)
discuss high--velocity ($\ge 23,000$ \kms) Ca~II in flux spectra of
the peculiar SN~2000cx.  Both Kasen et~al. and Thomas et~al. conclude
that the Ca~II features formed in non--spherical geometry, probably in
clumps located in front of the photosphere.  The spectra of SN~1998aq
do not extend to the Ca~II infrared triplet (and barely extend to the
high--velocity side of the Ca~II H\&K feature) so we have not
established the presence of high--velocity Ca~II features in
SN~1998aq.  In any case, the presence of high--velocity features in
early spectra of SNe~Ia evidently is not uncommon.  Further studies of
early--time polarization and flux spectra of SNe~Ia are needed to
clarify the composition, geometry, and origin of the high--velocity
matter.

Perhaps the most significant result of our analysis of SN~1998aq is
the evidence for C~II lines.  The possible presence of C~II in early
spectra of SNe~Ia has been discussed before.  For instance, Mazzali
(2001) tentatively attributed absorptions in a $-14$ day spectrum of
SN~1990N to C~II \lam6580 and \lam7234 forming near the photospheric
velocity of about 16,000 \kms.  The features that we attribute to
\lam6580 and \lam7234 in SN~1998aq are conspicuous in our figures
because of we plot $f_\nu/\nu$ to raise the red end of the spectra.
If C~II is present in SN~1998aq then it also is present in at least
some of the other SNe~Ia that have been observed at early times; e.g.,
similar features can be seen in spectra of SN~1994D (Fig.~4) and
SN~1996X (Salvo et~al. 2001).

The presence of C~II would have important implications for SN~Ia
explosion models.  In the synthetic spectrum of Fig. 10, we use C~II
detached at 14,000 \kms, but in Figs.~18 and 19, C~II is undetached
and therefore forming down to the photospheric velocity of 11,000
\kms, and we have not established the {\sl absence} of carbon at even
lower velocities.  Published delayed--detonation models for normal
SNe~Ia have very little carbon below about 30,000 \kms.  Model W7, a
parameterized one--dimensional (1D) deflagration model, has unburned
carbon down to 14,000 \kms.  Recently computed 3D deflagrations
(Khokhlov 2000; Gamezo et~al. 2003) have carbon at {\sl all} ejection
velocities.  The spectroscopic appearance of such models has begun to
be explored (Baron et~al. 2003).  In the context of existing models,
the presence of C~II lines may be consistent with deflagrations but
not with delayed--detonations.  However, the spectra of 3D
deflagration models appear to be inconsistent with SN~Ia spectra in
other respects (Thomas et~al. 2002; Gamezo et~al. 2003), while
delayed--detonations have been found to be consistent with certain
unusual SNe~Ia (Lentz et~al. 2001a) and in some respects with normal
SNe~Ia (Wheeler et~al 1998).  Much further work on comparing synthetic
spectra, both parameterized and detailed, with observed spectra of
SNe~Ia is needed to guide us towards satisfactory explosion models.

\bigskip

This work has been supported by NSF grants AST-9986965 and
AST-0204771 and NASA grant NAG5-12127.

\clearpage

\begin{figure}
\includegraphics[width=.8\textwidth]{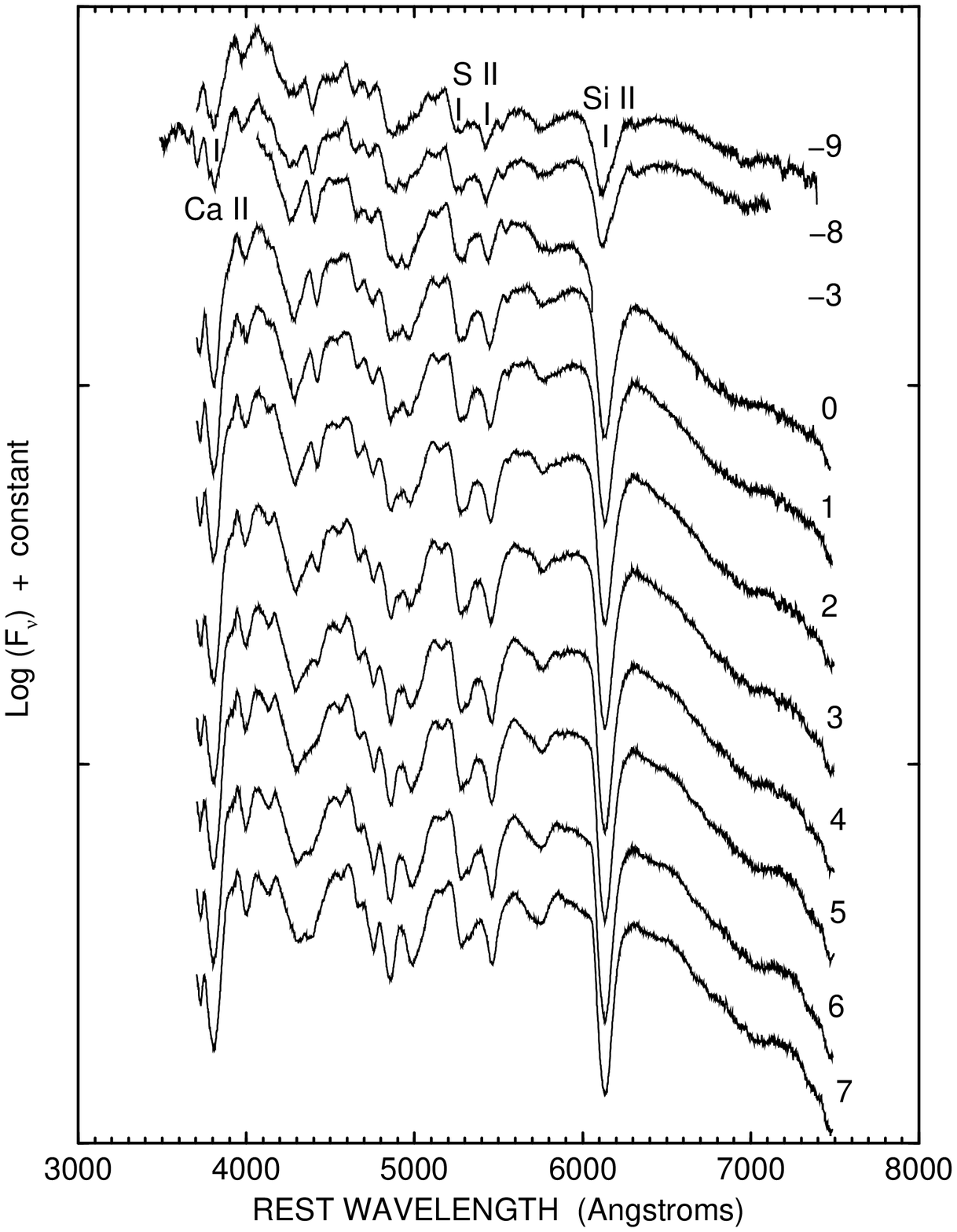}
\caption{Spectra of SN~1998aq.  Epochs are in days with respect to the
date of maximum brightness in the $B$ band, 1998 April~27.  The
vertical displacement is arbitrary.  All spectra shown in this paper
have been corrected for the redshift of NGC~3982, $z=0.003699$. No
correction for interstellar reddening has been applied.}
\end{figure}

\begin{figure}
\includegraphics[width=.8\textwidth]{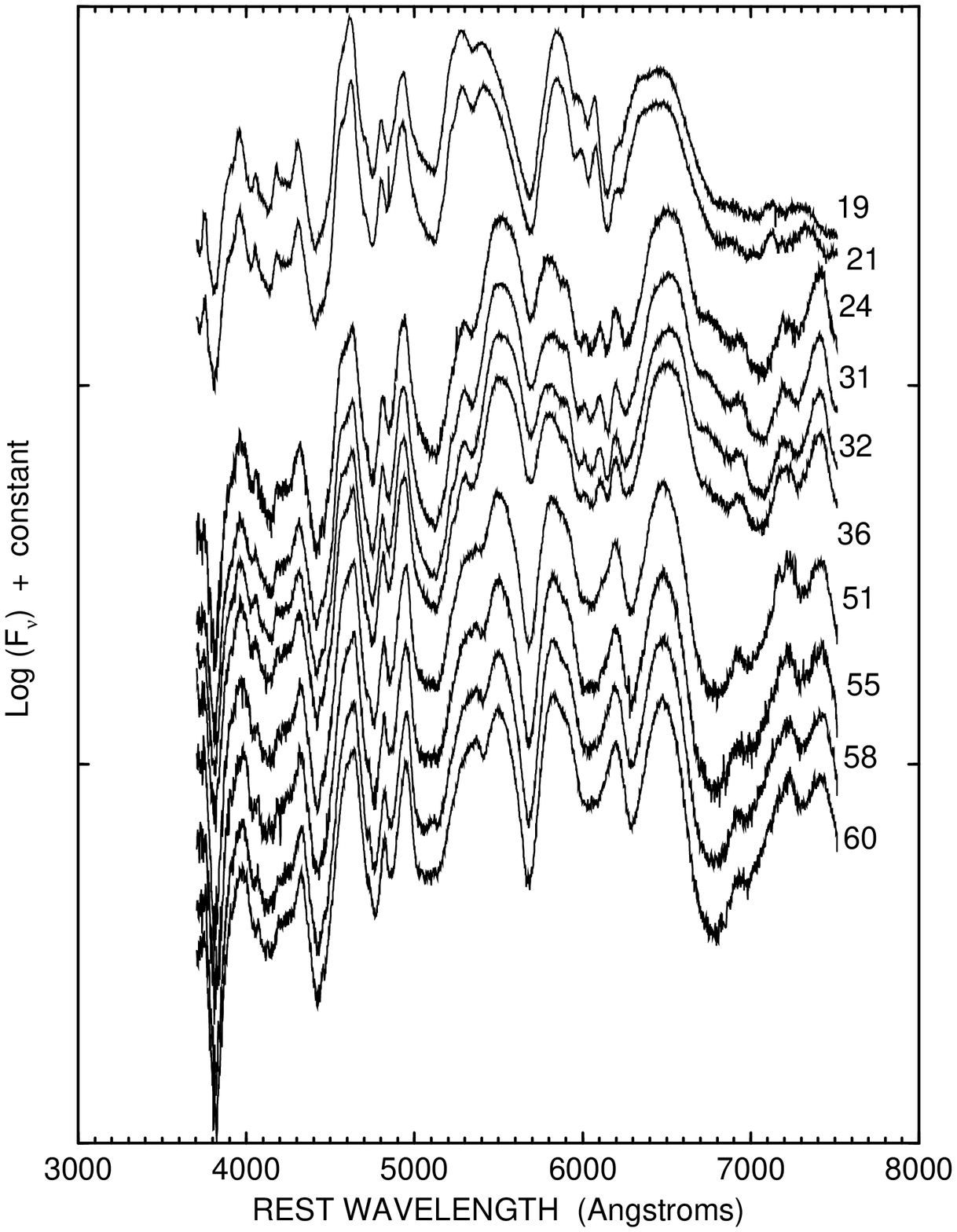}
\caption{Spectra of SN~1998aq.}
\end{figure}

\begin{figure}
\includegraphics[width=.8\textwidth]{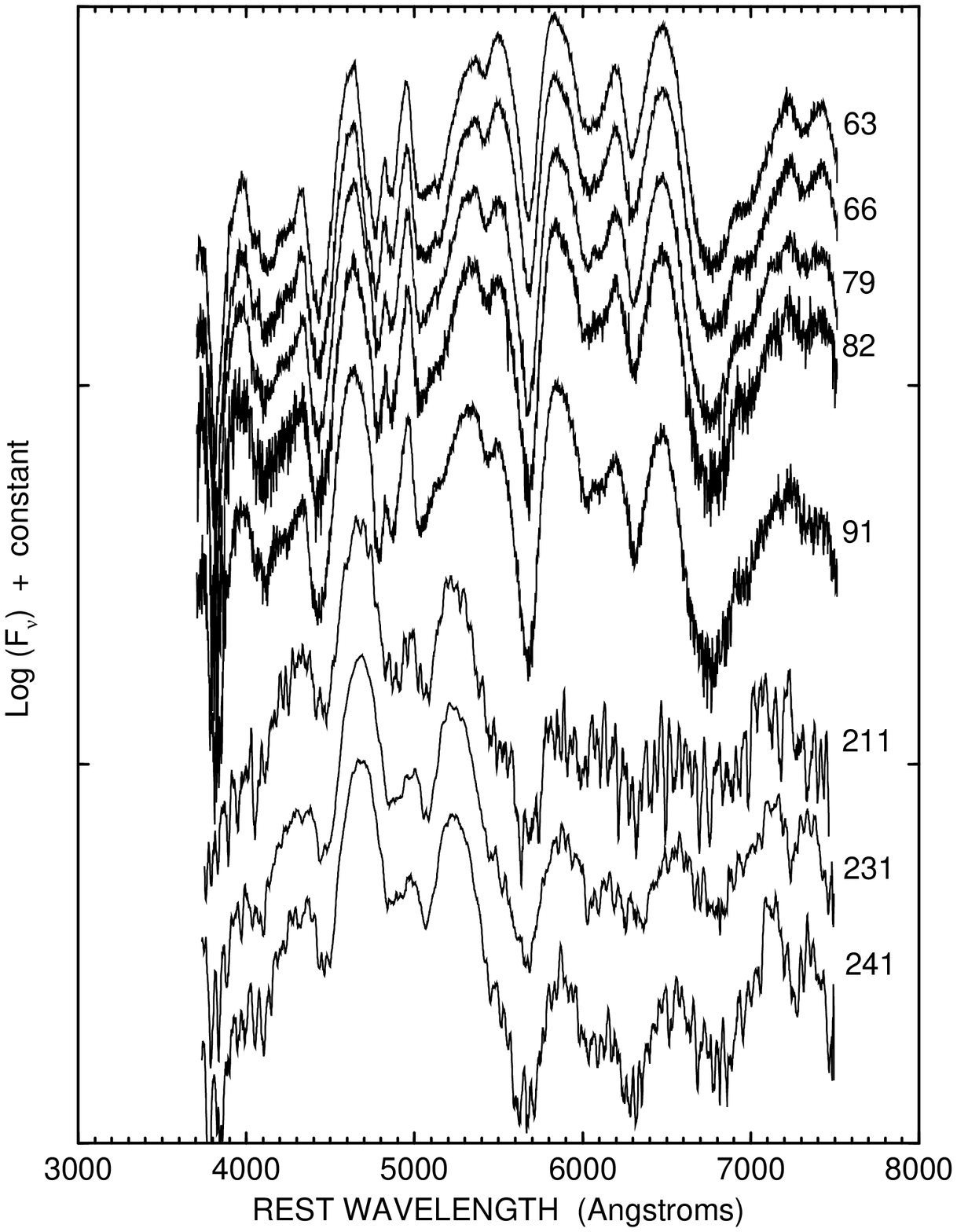}
\caption{Spectra of SN~1998aq.}
\end{figure}

\begin{figure}
\includegraphics[width=.9\textwidth]{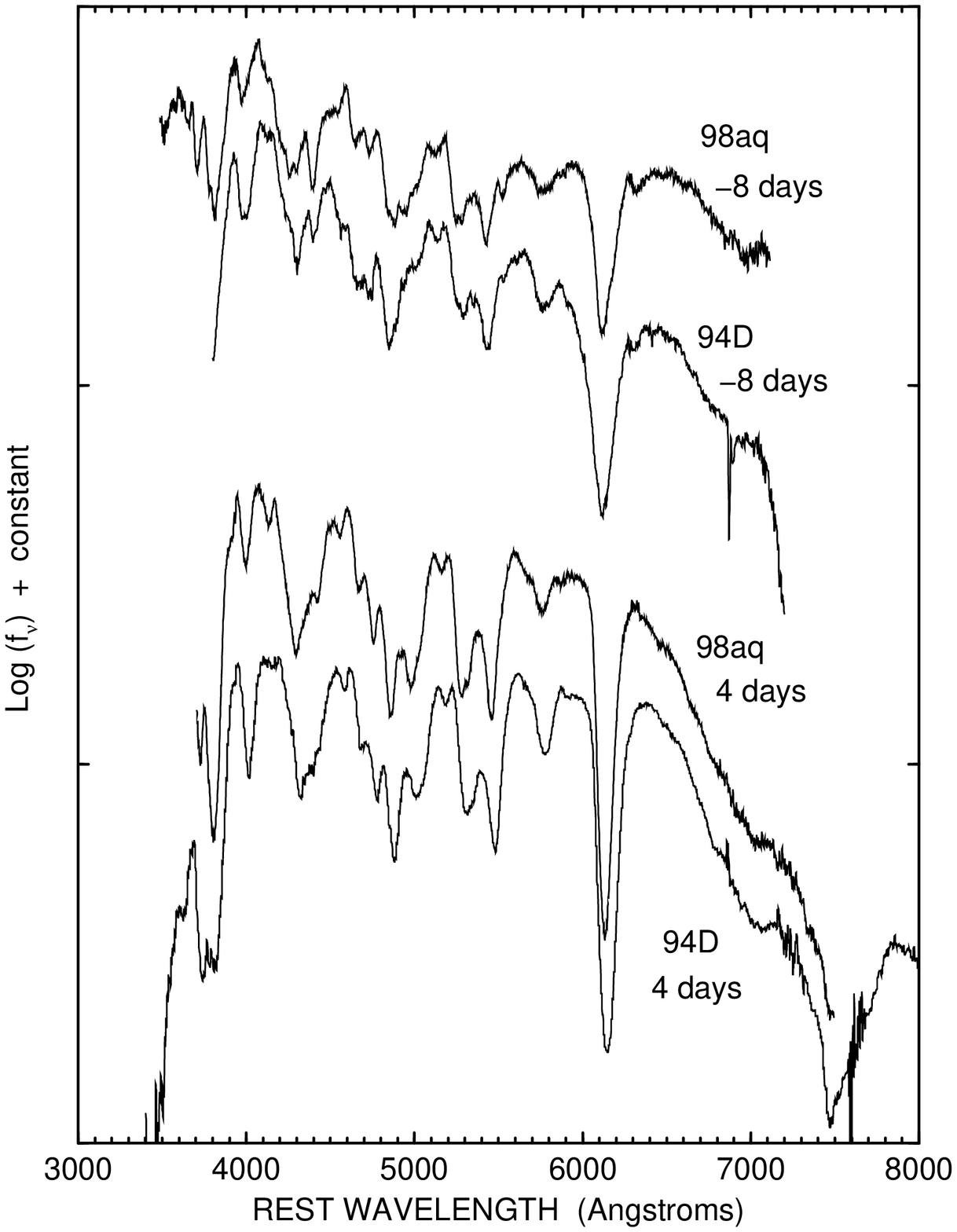}
\caption{Comparisons of spectra of SNe~1998aq and 1994D.}
\end{figure}

\begin{figure}
\includegraphics[width=.9\textwidth]{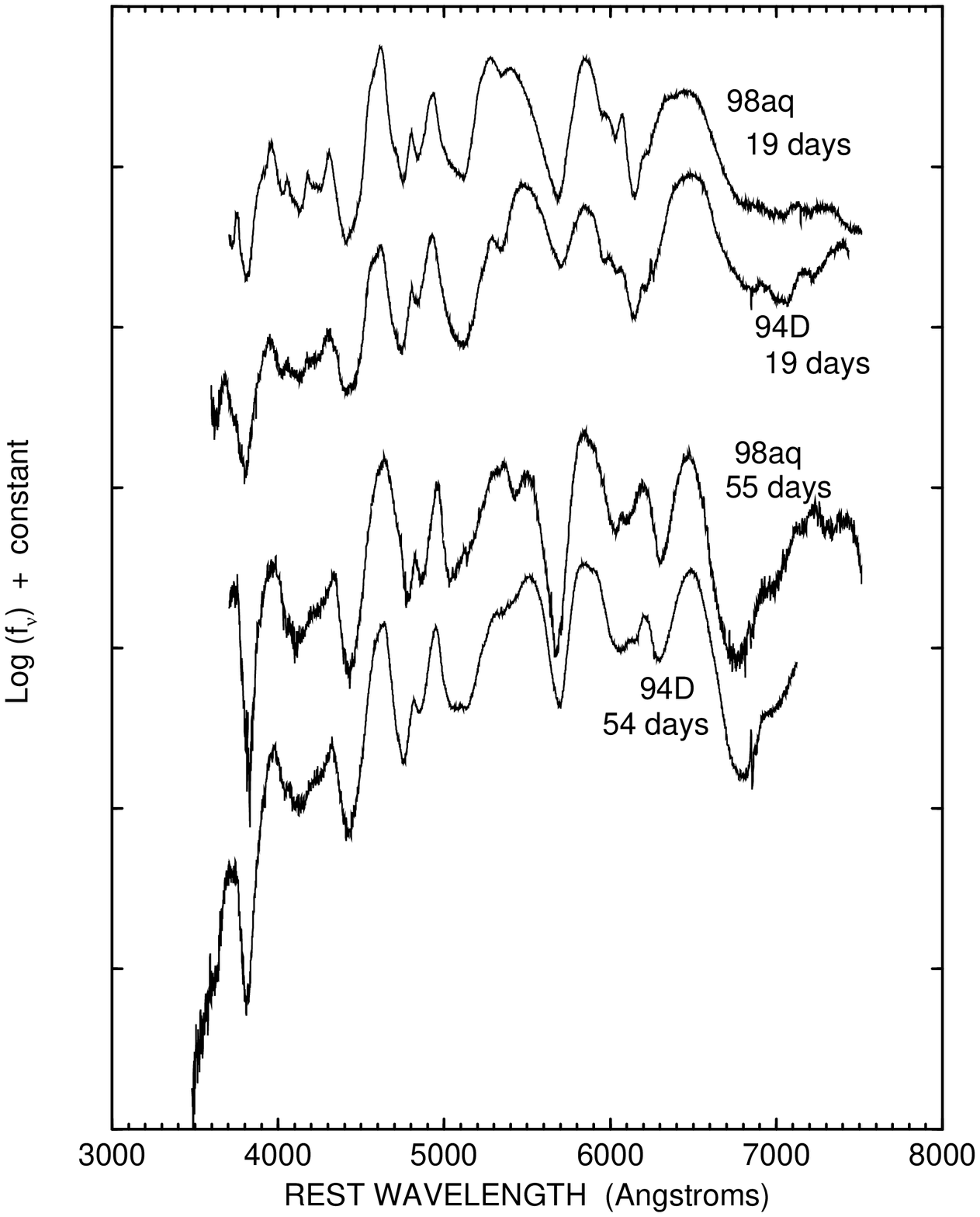}
\caption{Comparisons of spectra of SNe~1998aq and 1994D.}
\end{figure}

\clearpage

\begin{figure}
\includegraphics[width=.9\textwidth]{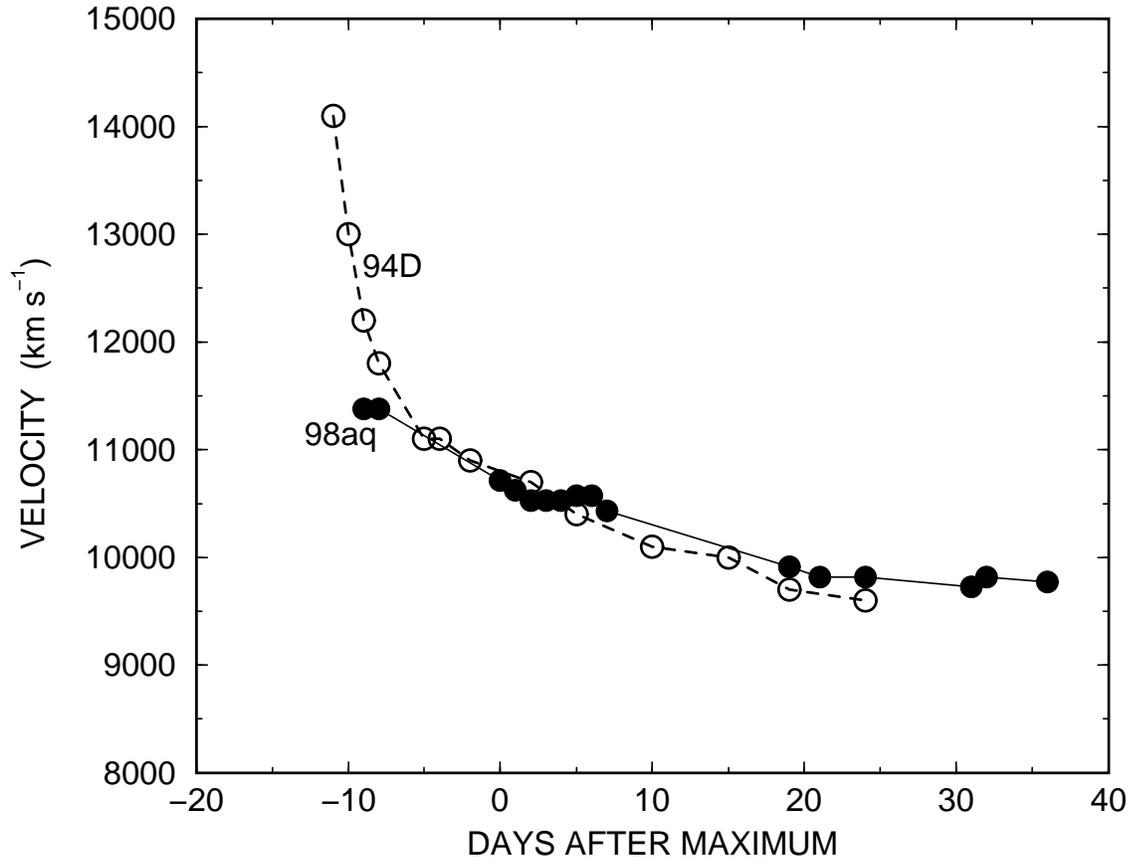}
\caption{Comparison of Si~II blueshifts in SNe~1998aq and 1994D.}
\end{figure}

\begin{figure}
\includegraphics[width=.9\textwidth]{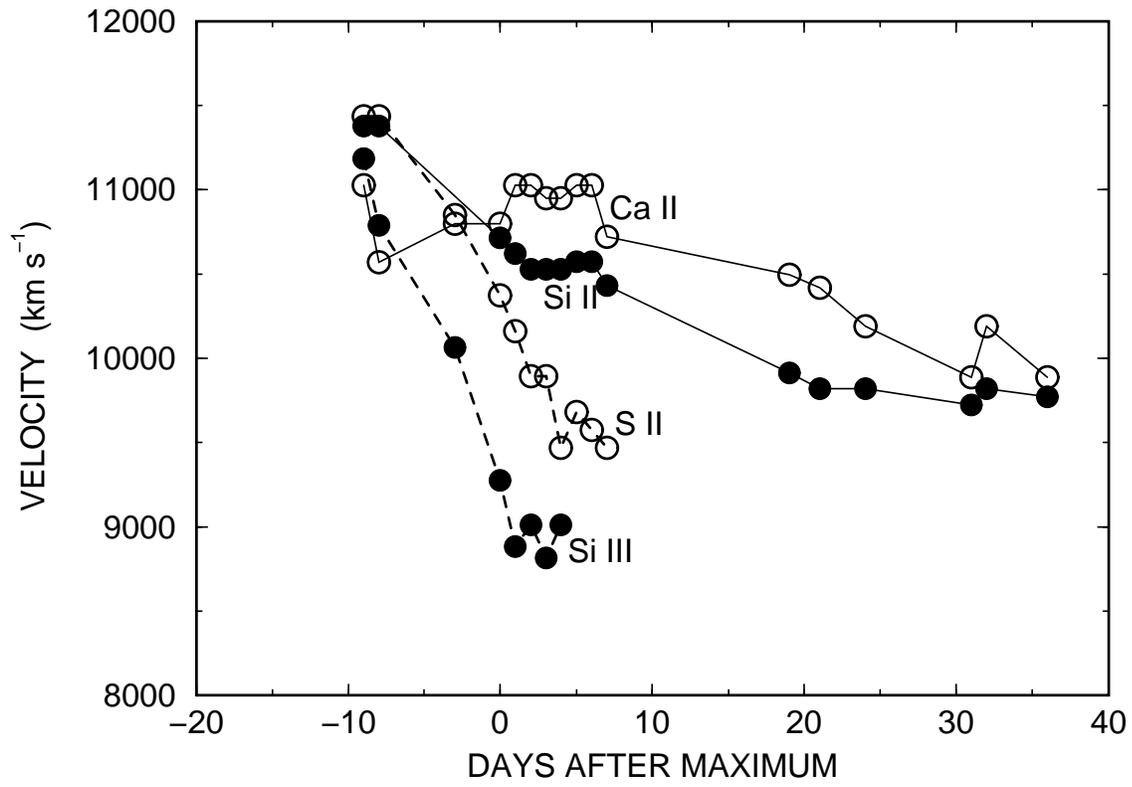}
\caption{Blueshifts of four absorption features in SN~1998aq.}
\end{figure}

\begin{figure}
\includegraphics[width=.9\textwidth]{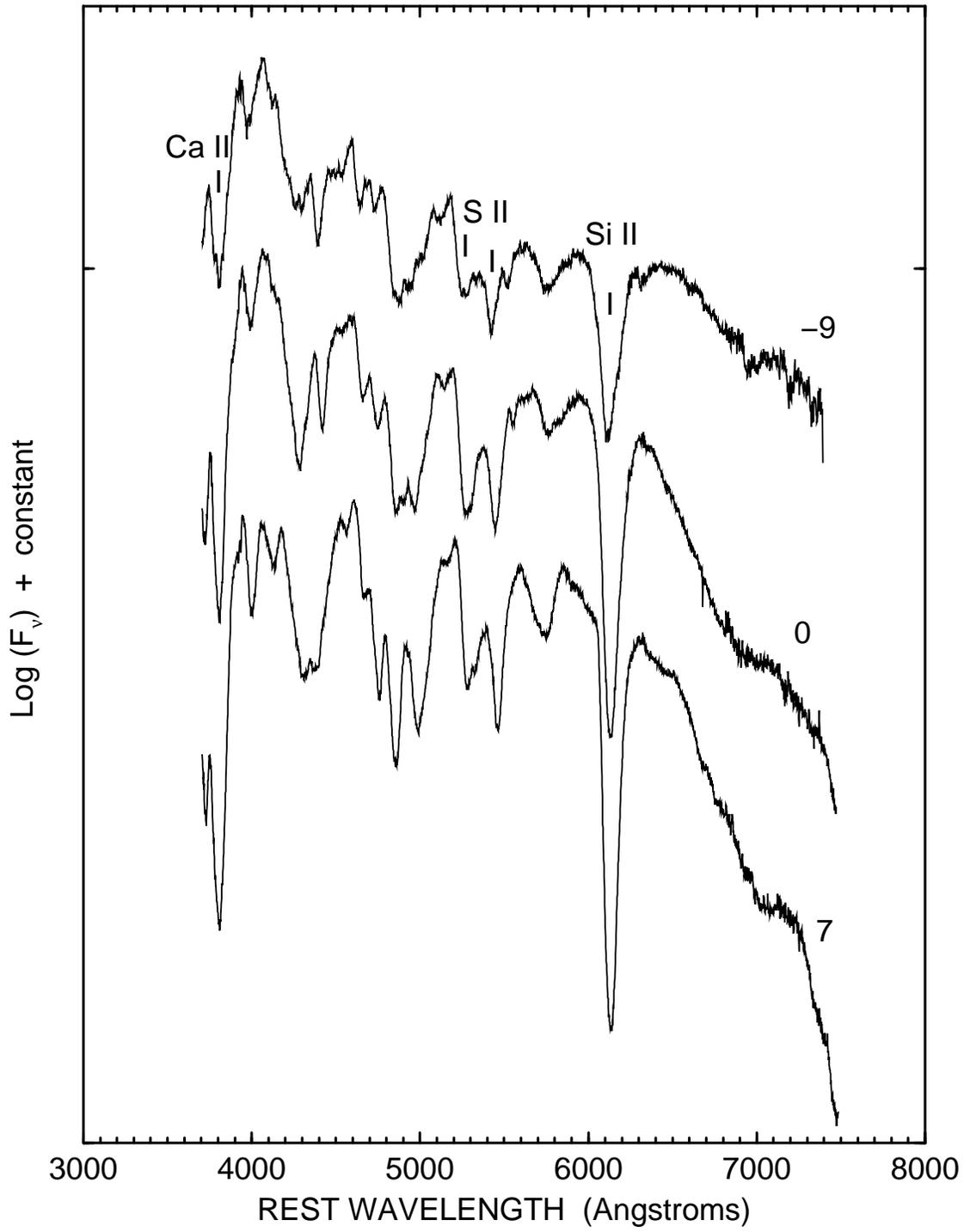}
\caption{Three early spectra of SN~1998aq.}
\end{figure}

\begin{figure}
\includegraphics[width=.9\textwidth]{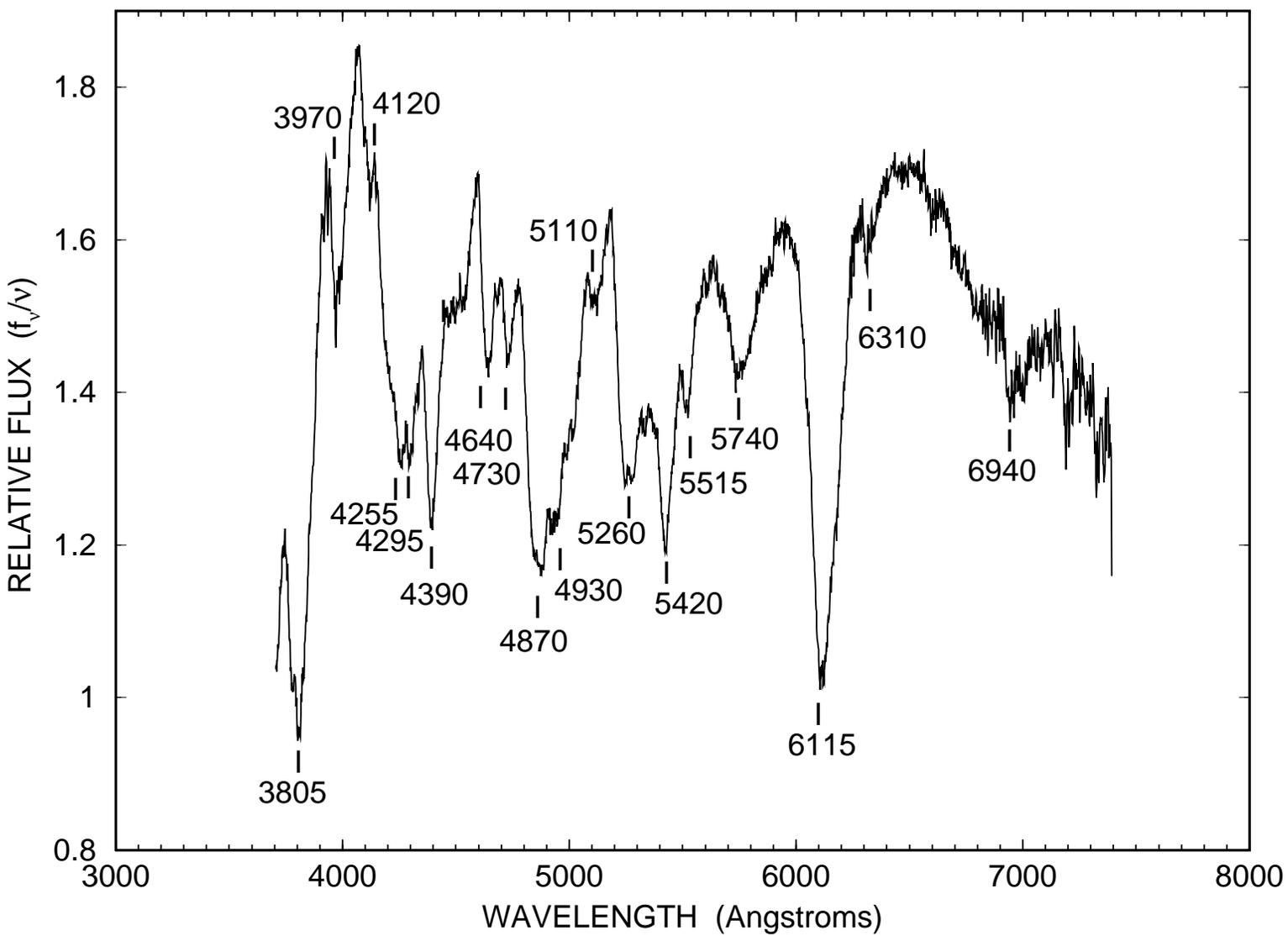}
\caption{Absorption features in the $-9$ day spectrum of
  SN~1998aq. Note that in this and subsequent figures the vertical
  axis is $f_\nu/\nu$.}
\end{figure}

\begin{figure}
\includegraphics[width=.9\textwidth]{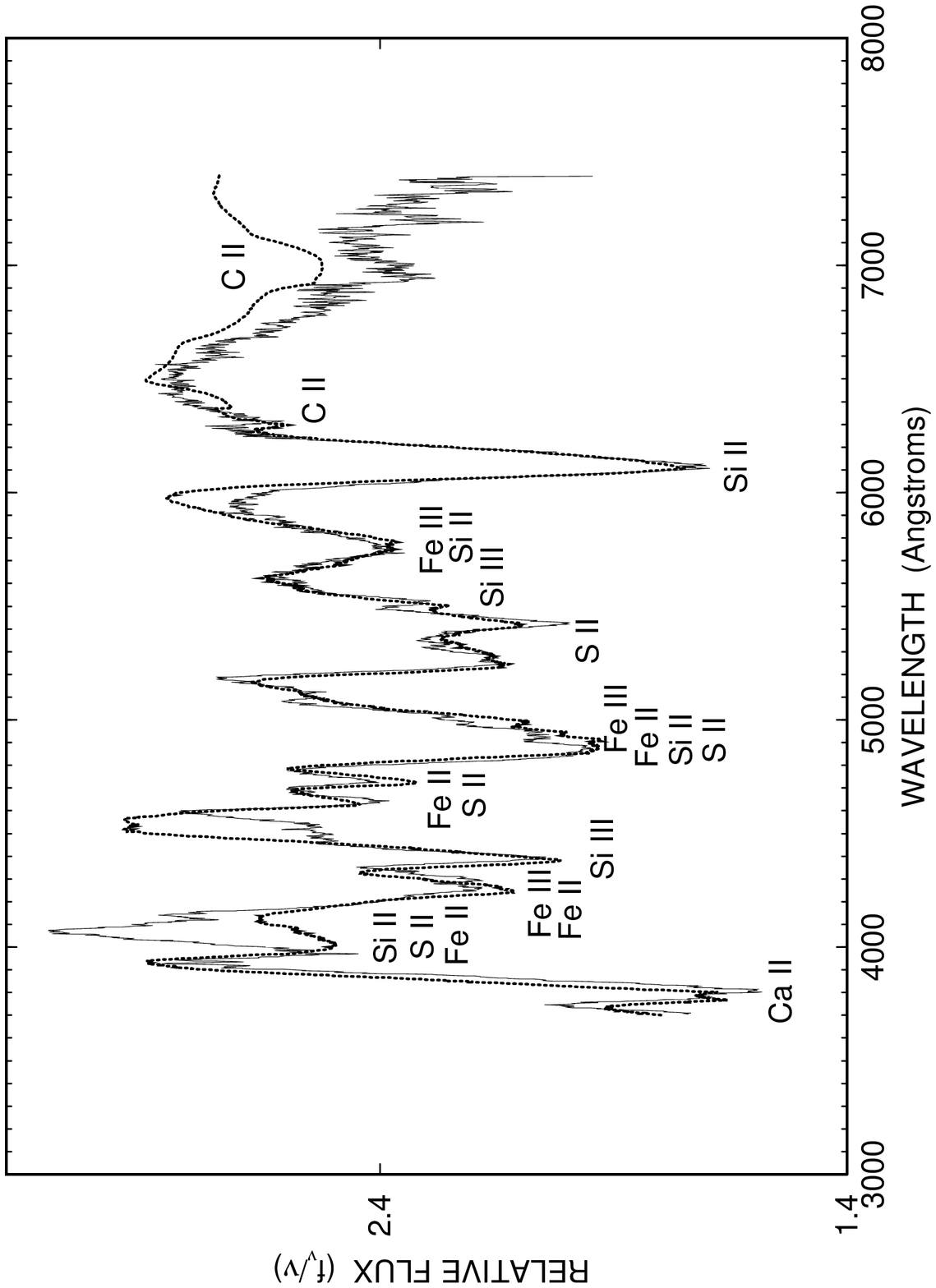}
\caption{Comparison of the $-9$ day spectrum of SN~1998aq with a {\bf
  Synow} synthetic spectrum that has $v_{phot}=13,000$ \kms,
  T$_{bb}=18,000$, and contains lines of seven ions.}
\end{figure}

\begin{figure}
\includegraphics[width=.9\textwidth]{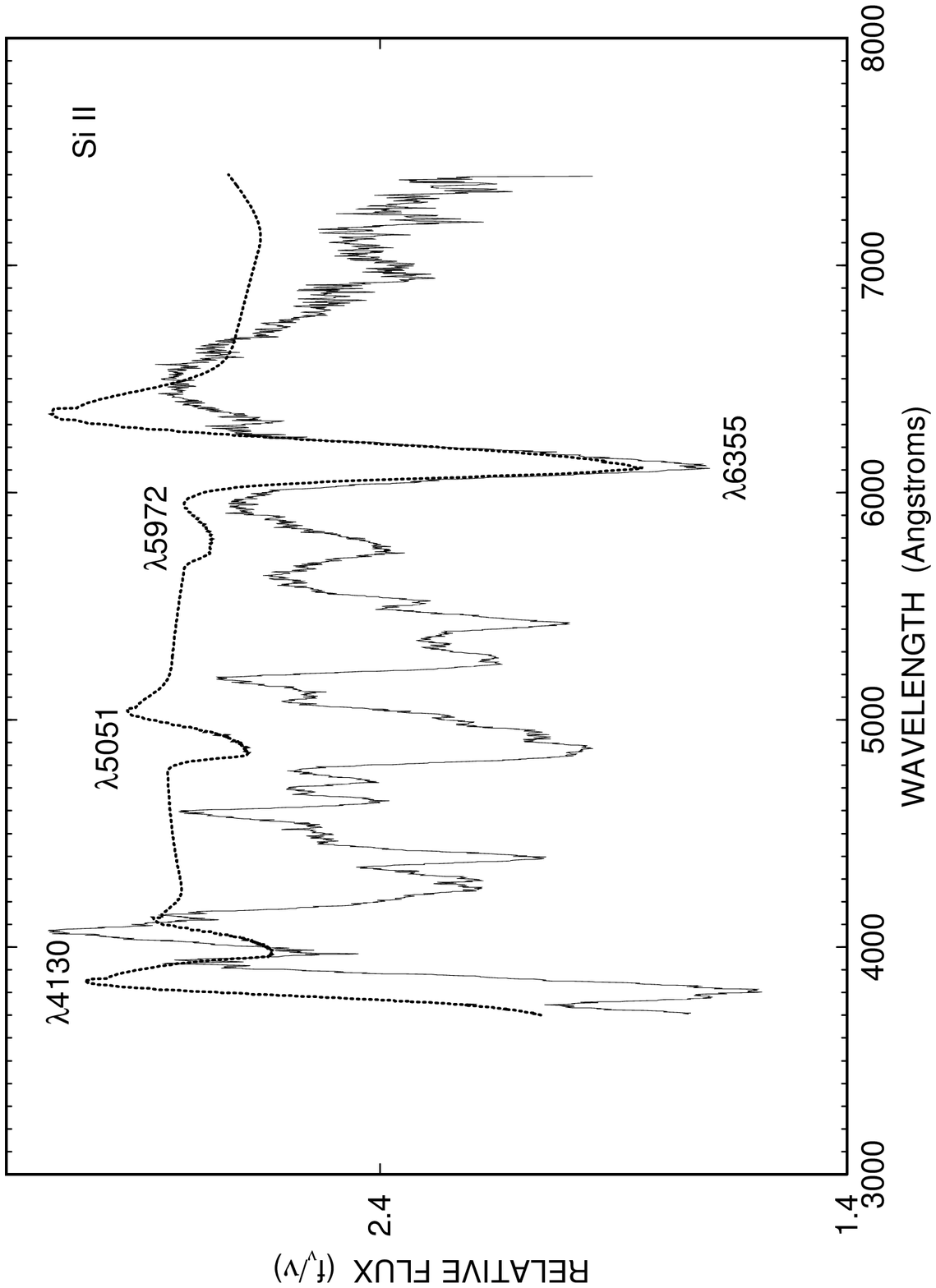}
\caption{Like Fig.~10 but with only lines of Si~II in the synthetic
  spectrum.}
\end{figure}

\clearpage

\begin{figure}
\includegraphics[width=.9\textwidth]{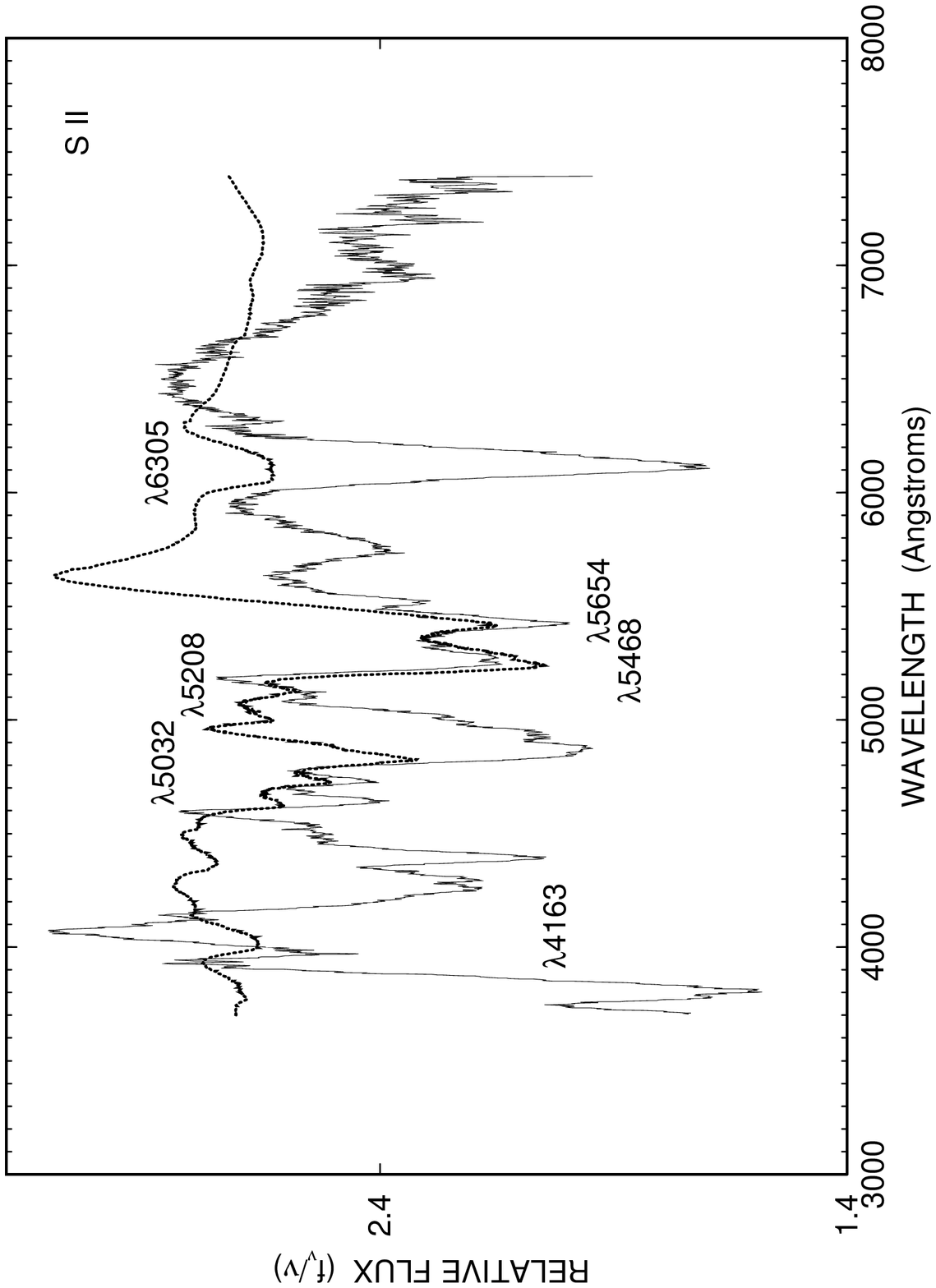}
\caption{Like Fig.~10 but with only lines of S~II in the synthetic
  spectrum.}
\end{figure}

\begin{figure}
\includegraphics[width=.9\textwidth]{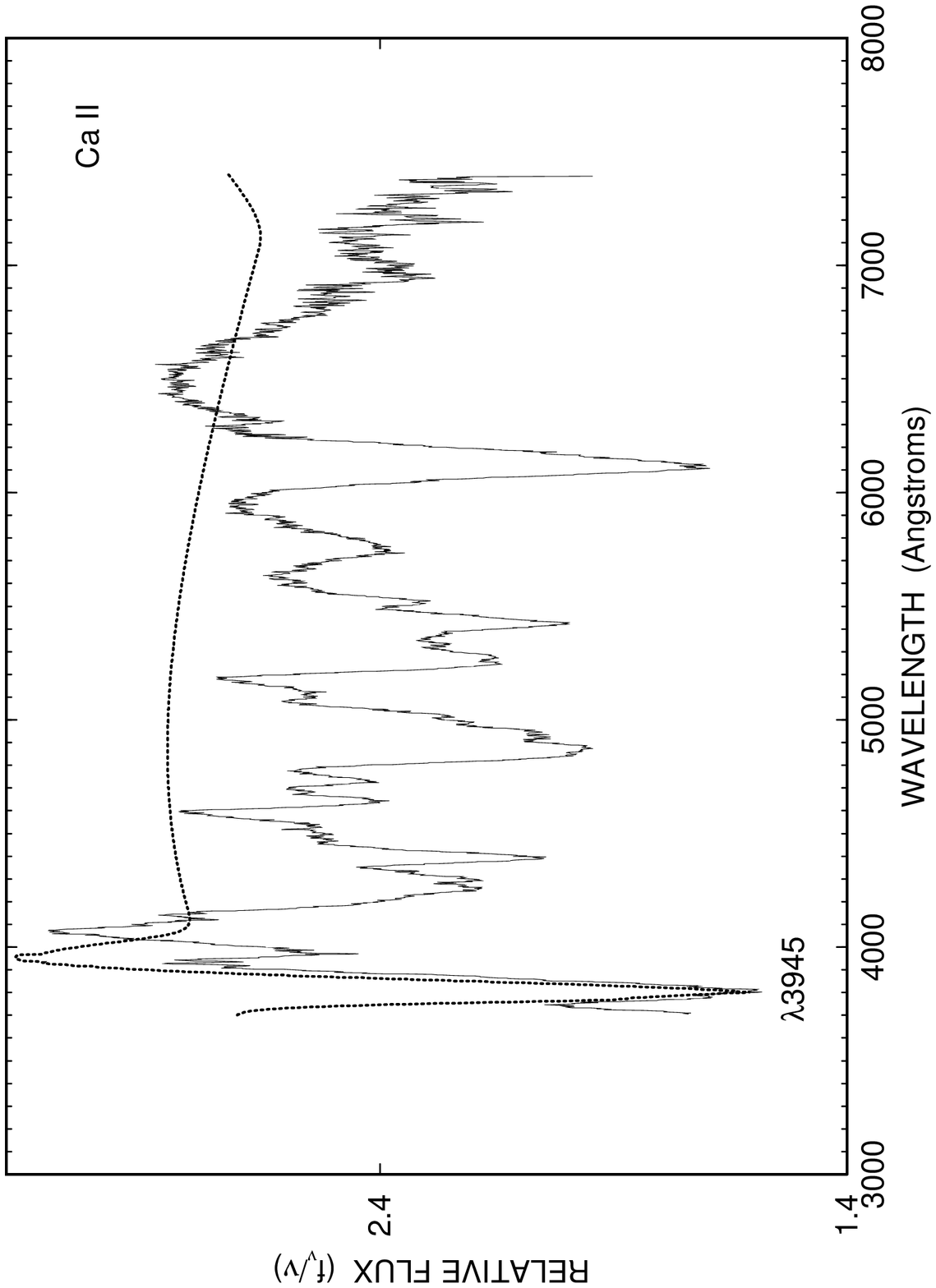}
\caption{Like Fig.~10 but with only lines of Ca~II in the synthetic
  spectrum.}
\end{figure}

\begin{figure}
\includegraphics[width=.9\textwidth]{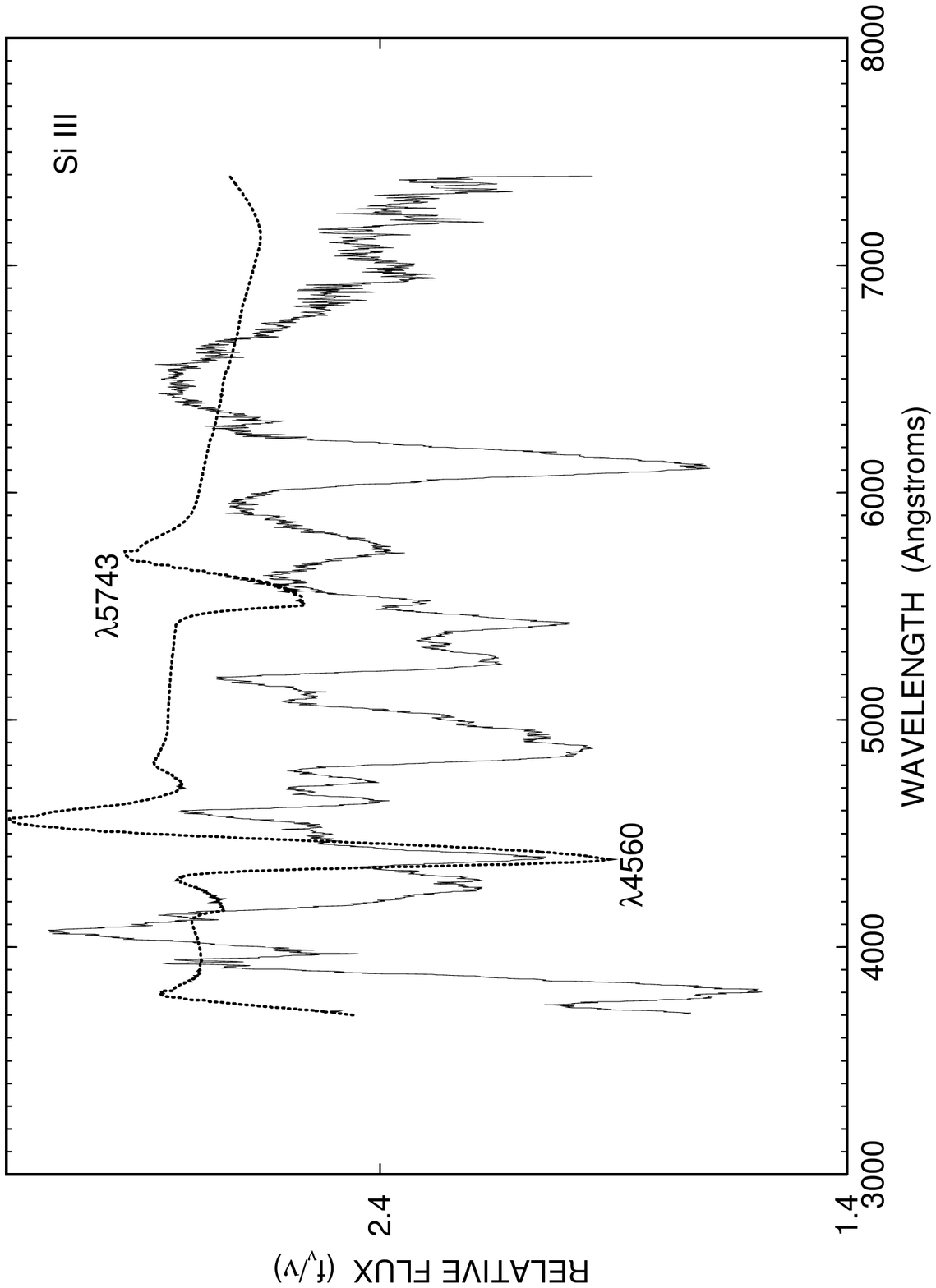}
\caption{Like Fig.~10 but with only lines of Si~III in the synthetic
  spectrum.}
\end{figure}

\begin{figure}
\includegraphics[width=.9\textwidth]{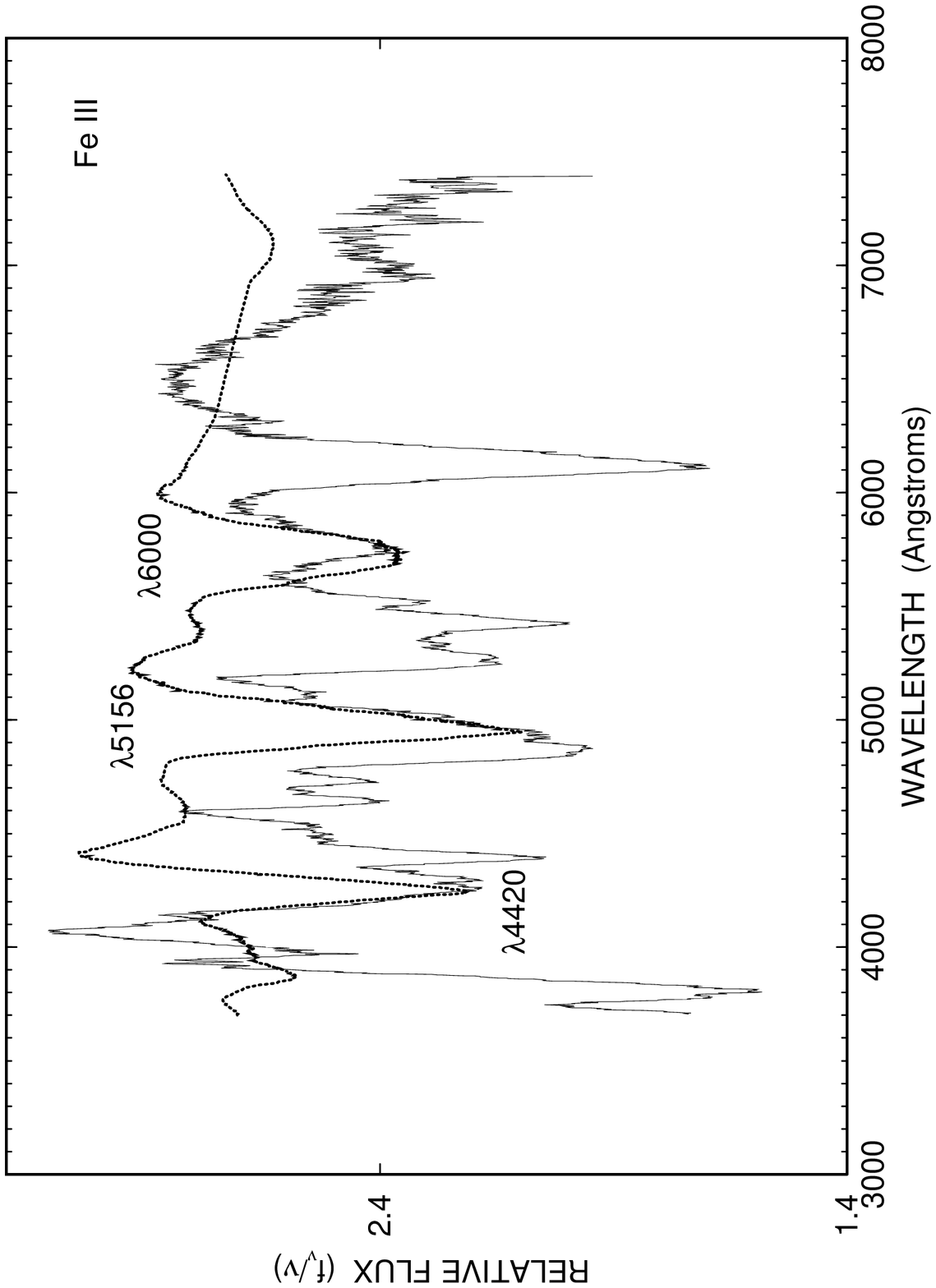}
\caption{Like Fig.~10 but with only lines of Fe~III in the synthetic
  spectrum.}
\end{figure}

\begin{figure}
\includegraphics[width=.9\textwidth]{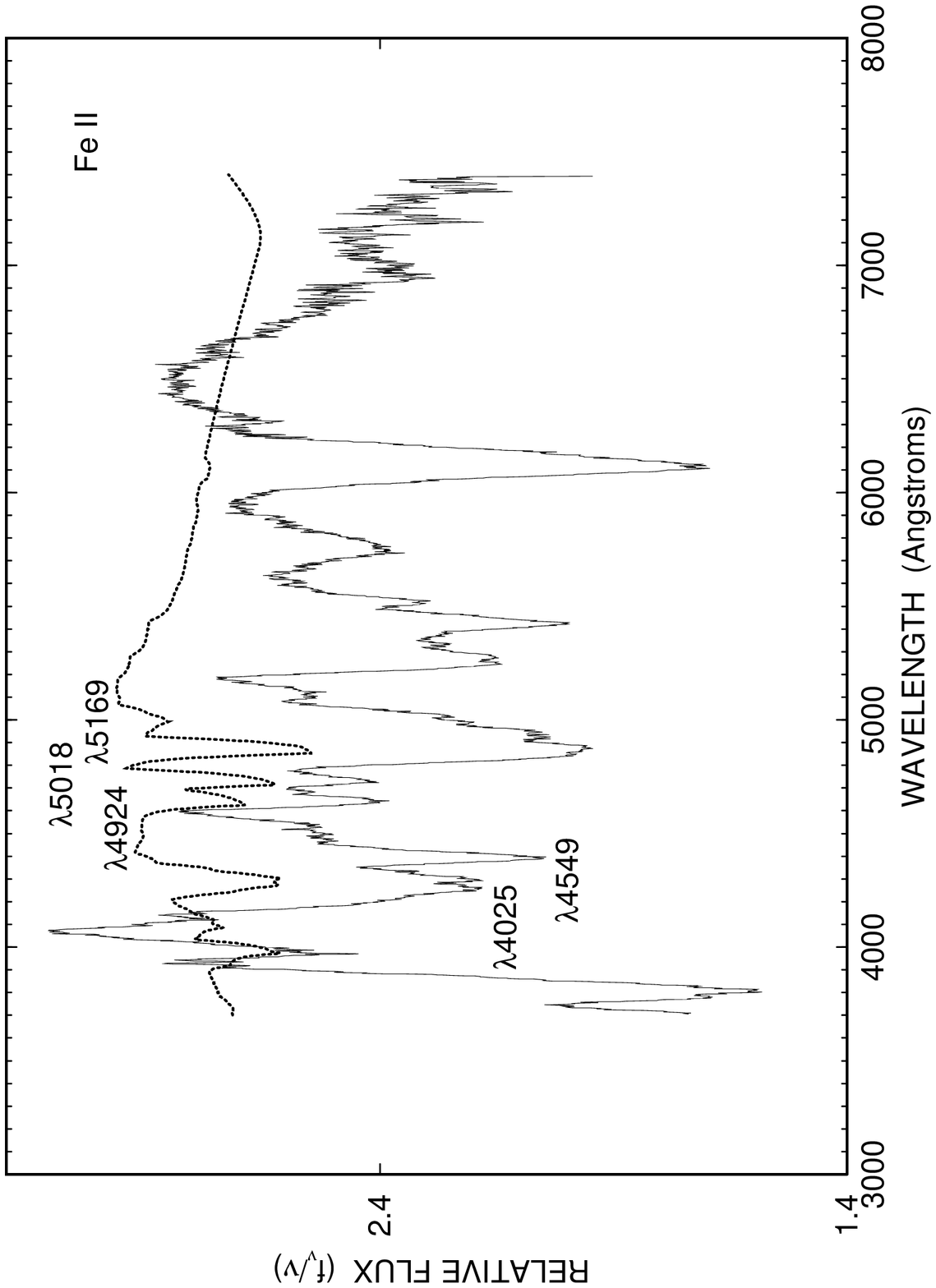}
\caption{Like Fig.~10 but with only lines of Fe~II in the synthetic
  spectrum.}
\end{figure}

\clearpage

\begin{figure}
\includegraphics[width=.9\textwidth]{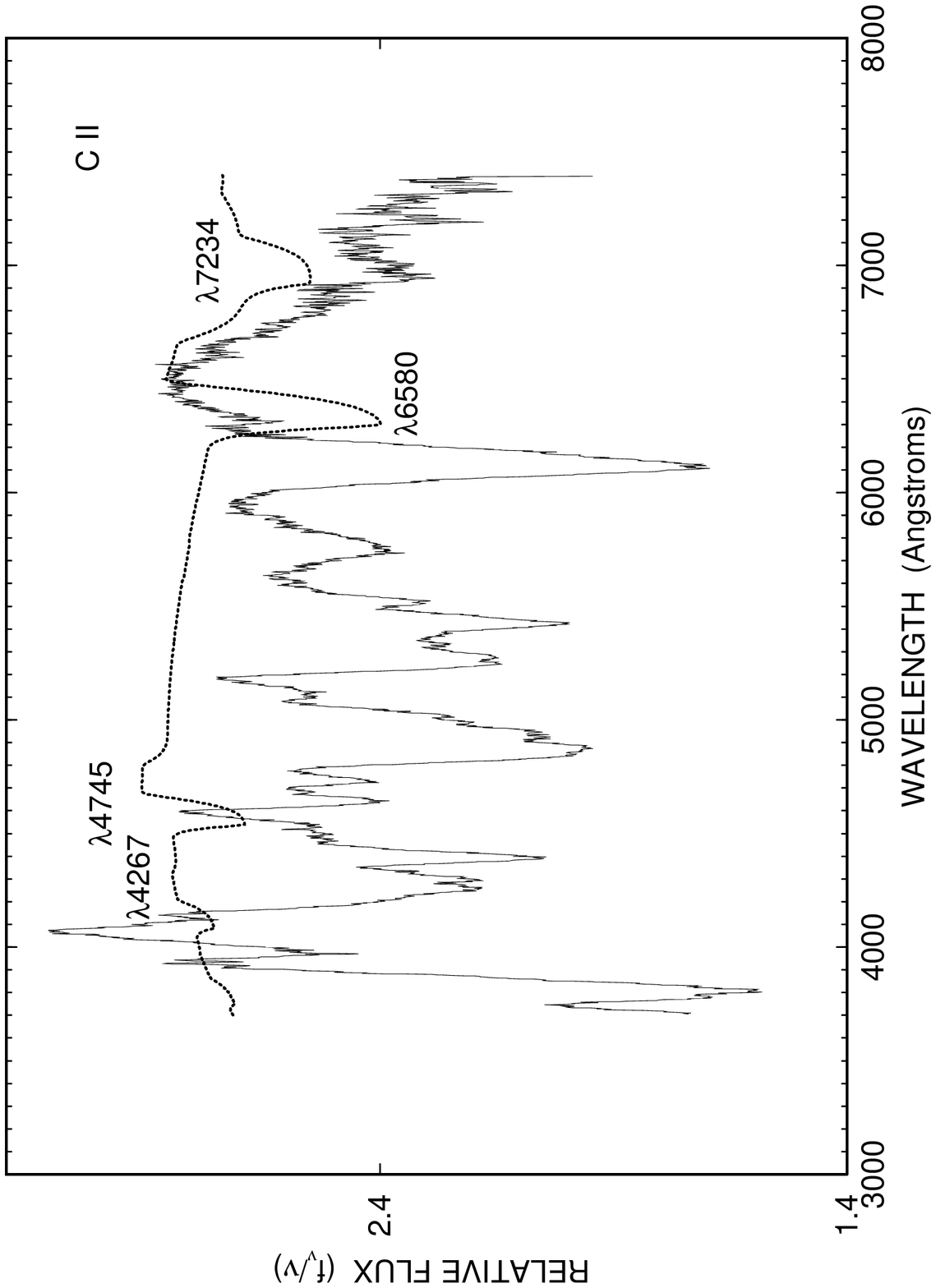}
\caption{Like Fig.~10 but with only lines of C~II in the synthetic
  spectrum.}
\end{figure}

\begin{figure}
\includegraphics[width=.9\textwidth]{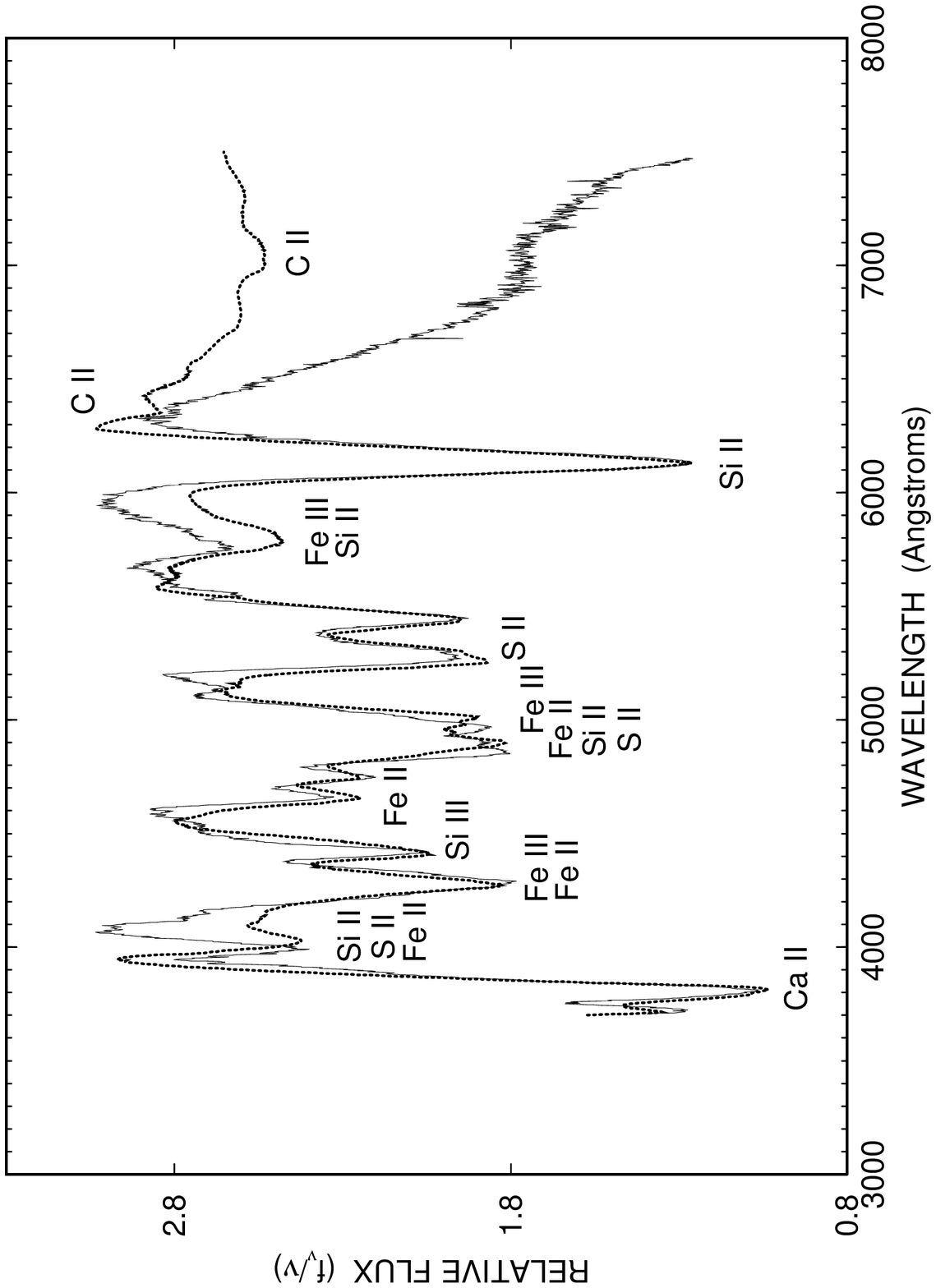}
\caption{Comparison of the $0$ day spectrum of SN~1998aq with a {\bf
  Synow} synthetic spectrum that has $v_{phot}=11,000$ \kms,
  T$_{bb}=16,000$, and contains lines of seven ions.}
\end{figure}

\begin{figure}
\includegraphics[width=.9\textwidth]{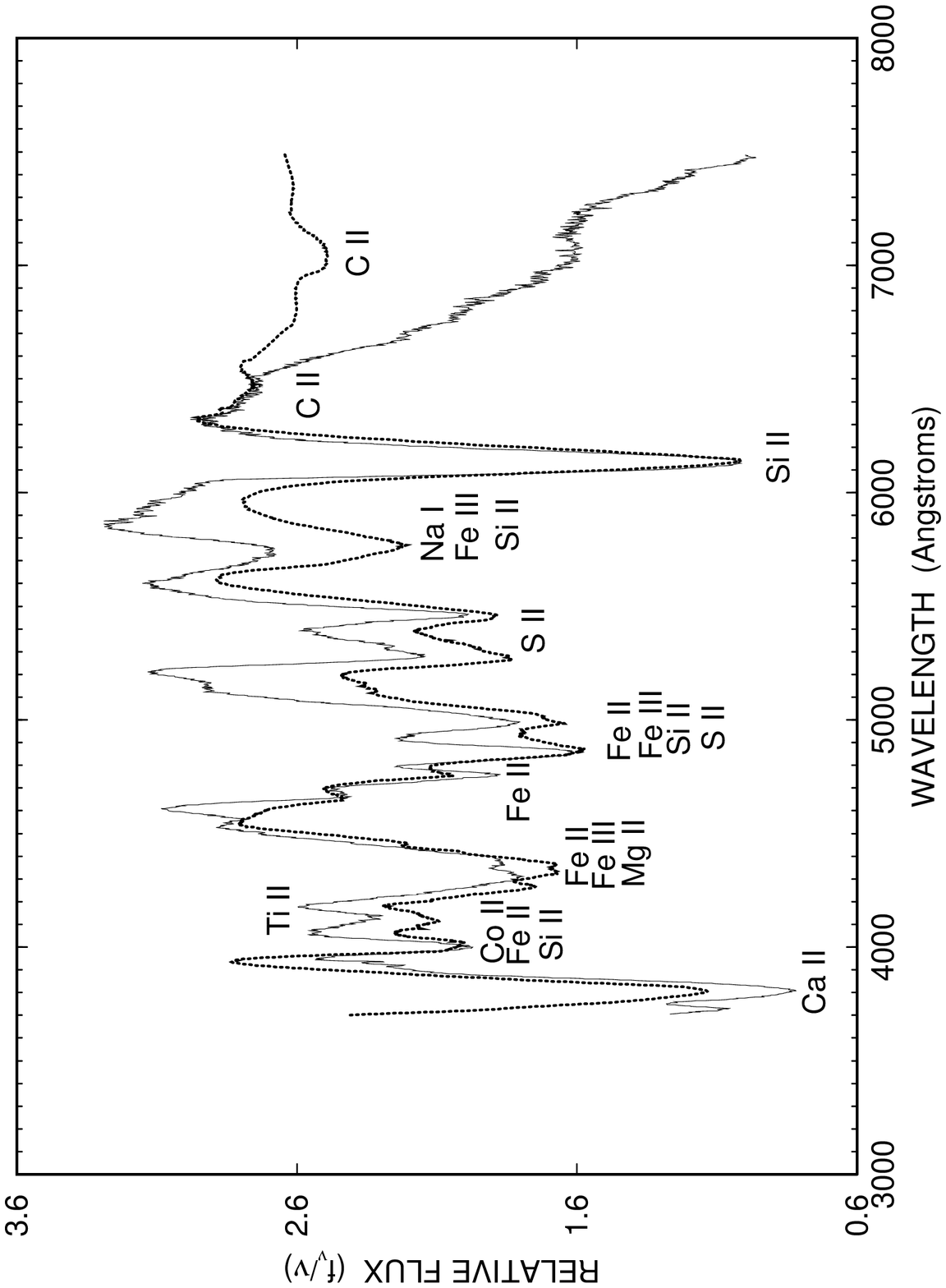}
\caption{Comparison of the $7$ day spectrum of SN~1998aq with a {\bf
  Synow} synthetic spectrum that has $v_{phot}=11,000$ \kms,
  T$_{bb}=16,000$, and contains lines of ten ions.}
\end{figure}

\begin{figure}
\includegraphics[width=.9\textwidth]{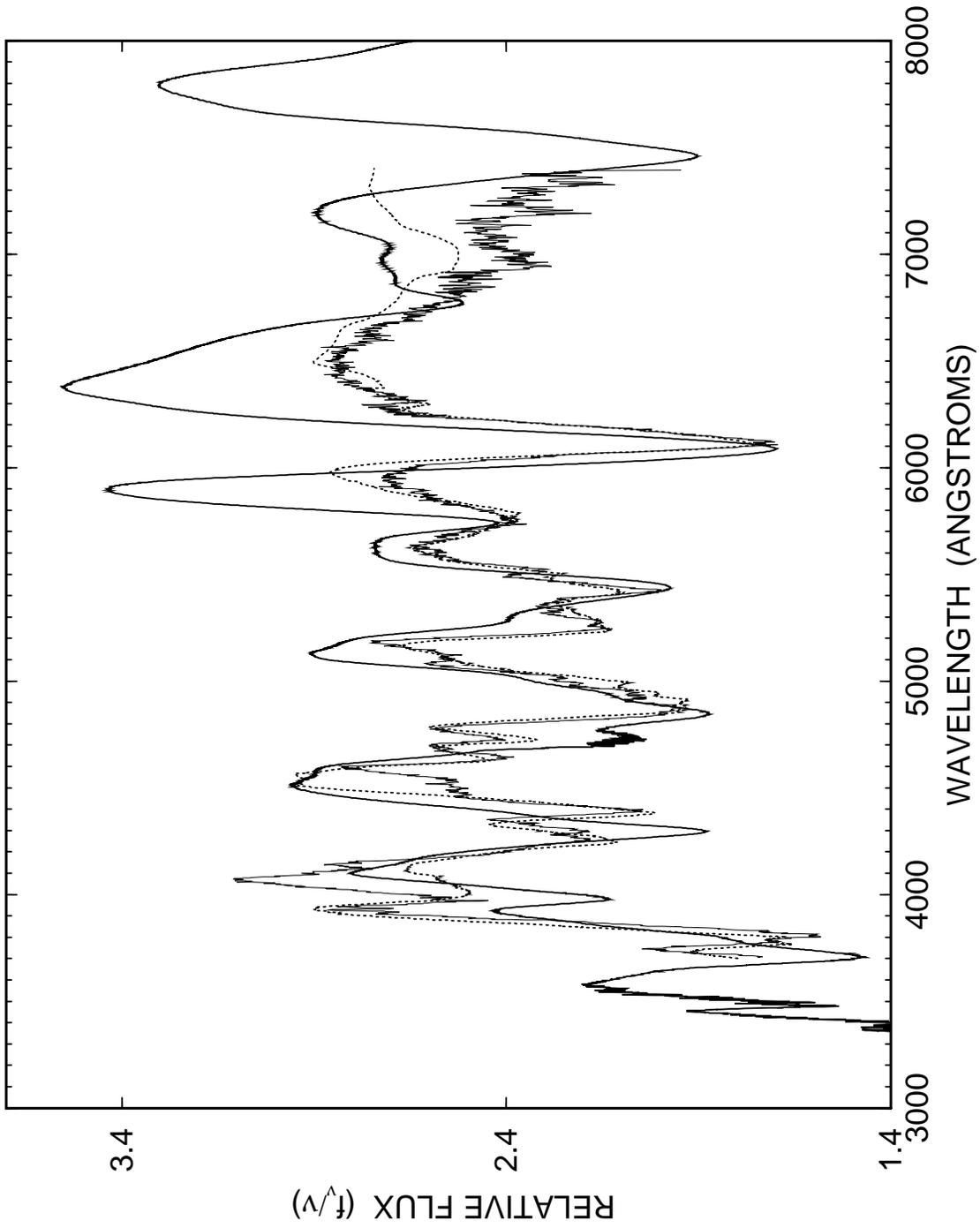}
\caption{Like Fig.10, but including a \texttt{PHOENIX} synthetic
  spectrum (smooth solid line) for model~W7 at 11 days after
  explosion, from Lentz et~al. (2001).}
\end{figure}

\clearpage

\begin{deluxetable}{lcccccccccccccl}
\tabletypesize{\scriptsize}
\rotate
\tablewidth{0pt}
\tablecaption{JOURNAL OF OBSERVATIONS}
\tablehead{\colhead{} & 
\colhead{UT Date} &
\colhead{Julian Day} &
\colhead{Tel.} &
\colhead{Range}  &
\colhead{Res.} &
\colhead{P.A.} &
\colhead{Par.} &
\colhead{Air.} & 
\colhead{Flux Std.} &
\colhead{Sky} &
\colhead{See.} &
\colhead{Slit} &
\colhead{Exp.} &
\colhead{Observer(s)} \\
\colhead{} &
\colhead{} &
\colhead{} &
\colhead{} &
\colhead{(\AA)} &
\colhead{(\AA)} &
\colhead{($^\circ$)} &
\colhead{($^\circ$)} &
\colhead{} &
\colhead{} &
\colhead{} &
\colhead{($^{\prime\prime}$)} &
\colhead{($^{\prime\prime}$)} &
\colhead{(s)} & 
\colhead{} }
\startdata
 & 1998-04-18.28 & 2450921.78 & FLWO & 3720.00-7420.5 & 7.0 &   90.00 &  149.67 & 1.12 &    Feige34 &      clear &    1-2 & 3 &  900 & P Berlind / M Calk \\
 & 1998-04-19.26 & 2450922.76 & FLWO &      3270-7140 & 7.0 &   90.00 &  163.01 & 1.10 &   Feige110 &      clear &      2 & 3 &  900 & P Berlind / M Calk \\
 & 1998-04-24.36 & 2450927.86 & FLWO & 4080.00-6080.2 & 7.0 &  110.00 &  104.63 & 1.38 &    Feige34 &      clear &      2 & 2 & 1200 &   Kannappan/Lepore \\
 & 1998-04-27.29 & 2450930.79 & FLWO &      3720-7500 & 7.0 &  110.00 &  130.62 & 1.17 &    Feige34 &      clear &      2 & 2 &  600 &          P Berlind \\
 & 1998-04-28.26 & 2450931.76 & FLWO & 3720.00-7510.5 & 7.0 &   90.00 &  148.58 & 1.12 &    Feige34 & mostly cle &      2 & 3 &  600 &          P Berlind \\
 & 1998-04-29.26 & 2450932.76 & FLWO &      3720-7521 & 7.0 &   90.00 &  145.02 & 1.13 &    Feige34 & mostly cle &     2+ & 3 &  600 &          P Berlind \\
 & 1998-04-30.24 & 2450933.74 & FLWO &      3720-7521 & 7.0 &   90.00 &  152.80 & 1.11 &    Feige34 & mostly cle &     2+ & 3 &  600 &          M Calkins \\
 & 1998-05-01.25 & 2450934.75 & FLWO &      3720-7521 & 7.0 &   90.00 &  145.21 & 1.13 &    Feige34 &      clear &      3 & 3 &  600 &          M Calkins \\
 & 1998-05-02.23 & 2450935.73 & FLWO &      3720-7521 & 7.0 &   90.00 &  158.42 & 1.10 &    Feige34 &      clear &      3 & 3 &  600 &          M Calkins \\
 & 1998-05-03.24 & 2450936.74 & FLWO &      3720-7515 & 7.0 &   90.00 &  148.00 & 1.12 &    Feige34 & mostly cle &    1-2 & 3 &  600 &          P Berlind \\
 & 1998-05-04.26 & 2450937.76 & FLWO &      3720-7512 & 7.0 &   90.00 &  136.26 & 1.15 &    Feige34 & lots of ci &    1-2 & 3 &  600 &          P Berlind \\
 & 1998-05-16.17 & 2450949.67 & FLWO &    3720-7540.5 & 7.0 &   90.00 &  176.89 & 1.09 &    Feige34 & mostly cle &    1-2 & 3 &  600 &          P Berlind \\
 & 1998-05-18.20 & 2450951.70 & FLWO &    3720-7540.5 & 7.0 &   90.00 &  151.82 & 1.11 &    Feige34 & scattered  &      1 & 3 &  600 &         M. Calkins \\
 & 1998-05-28.22 & 2450961.72 & FLWO &    3720-7540.5 & 7.0 &   91.00 &  122.30 & 1.21 &    Feige34 &      clear &      2 & 3 &  660 &          P Berlind \\
 & 1998-05-29.19 & 2450962.69 & FLWO &    3720-7540.5 & 7.0 &   91.00 &  136.41 & 1.15 &    Feige34 & cirrus to  &    2-3 & 3 &  660 &          P Berlind \\
 & 1998-05-31.20 & 2450964.70 & FLWO &    3720-7540.5 & 7.0 &   91.00 &  131.81 & 1.17 &    Feige34 & cirrus all &      2 & 3 &  900 &          M Calkins \\
 & 1998-06-02.23 & 2450966.73 & FLWO &    3720-7540.5 & 7.0 &   91.00 &  113.20 & 1.27 &  BDp284211 &      Clear &      3 & 3 &  600 &          M Calkins \\
 & 1998-06-17.17 & 2450981.67 & FLWO &    3720-7540.5 & 7.0 &   90.00 &  123.89 & 1.21 &  BDp284211 &      Clear &    1-2 & 3 &  900 &          B. Carter \\
 & 1998-06-21.18 & 2450985.68 & FLWO &    3720-7540.5 & 7.0 &   90.00 &  112.71 & 1.28 &  BDp284211 &      Clear &    1-2 & 3 &  900 &         M. Calkins \\
 & 1998-06-24.19 & 2450988.69 & FLWO &    3720-7540.5 & 7.0 &   90.00 &  106.38 & 1.35 &  BDp284211 &     cirrus &    1-2 & 3 &  900 &          P Berlind \\
 & 1998-06-26.19 & 2450990.69 & FLWO &    3720-7540.5 & 7.0 &   90.00 &  104.67 & 1.36 &  BDp284211 & hazy but c &    1-2 & 3 &  720 &  P Berlind/K Rines \\
 & 1998-06-29.17 & 2450993.67 & FLWO &    3720-7540.5 & 7.0 &   90.00 &  106.92 & 1.34 &  BDp284211 & hazy but c &    1-2 & 3 &  900 &            K Rines \\
 & 1998-07-02.17 & 2450996.67 & FLWO &    3720-7540.5 & 7.0 &   90.00 &  103.76 & 1.37 &  BDp284211 &     clear! &     2+ & 3 &  660 &          P Berlind \\
 & 1998-07-15.21 & 2451009.71 & FLWO &    3720-7540.5 & 7.0 &   90.00 &   82.10 & 1.92 &  BDp284211 &       hazy &      2 & 3 & 1200 &          P Berlind \\
 & 1998-07-18.18 & 2451012.68 & FLWO &    3720-7540.5 & 7.0 &   90.00 &   87.61 & 1.67 &  BDp284211 & major buil &        & 3 &  600 &          M Calkins \\
 & 1998-07-27.17 & 2451021.67 & FLWO &    3720-7540.5 & 7.0 &   90.00 &   82.81 & 1.86 &  BDp284211 & a few scat &      2 & 3 &  900 &          P Berlind \\
 & 1998-11-24.54 & 2451142.04 & FLWO &    3720-7540.5 & 7.0 &   90.00 & -114.52 & 1.20 &    Feige34 &      clear &     1+ & 3 & 1200 & K Rines, J. Huchra \\
 & 1998-11-24.54 & 2451142.04 & FLWO &    3720-7540.5 & 7.0 &   90.00 & -119.80 & 1.18 &    Feige34 &      clear &     1+ & 3 &  900 & K Rines, J. Huchra \\
 & 1998-12-14.53 & 2451162.03 & FLWO &    3720-7540.5 & 7.0 &   59.00 & -139.75 & 1.11 &    Feige34 &      clear &    1-2 & 3 & 1200 & M Calkins, A Mahda \\
 & 1998-12-14.55 & 2451162.05 & FLWO &    3720-7540.5 & 7.0 &   59.00 & -148.65 & 1.10 &    Feige34 &      clear &    1-2 & 3 & 1200 & M Calkins, A Mahda \\
 & 1998-12-24.53 & 2451172.03 & FLWO &    3720-7540.5 & 7.0 &   90.00 & -157.34 & 1.09 &    Feige34 &     cirrus &      2 & 3 & 1200 &          M Calkins \\
 & 1998-12-24.55 & 2451172.05 & FLWO &    3720-7540.5 & 7.0 &   90.00 & -167.64 & 1.09 &    Feige34 &     cirrus &      2 & 3 & 1200 &          M Calkins \\
\enddata
\end{deluxetable}
\clearpage

\begin{deluxetable}{crcrr}
\footnotesize
\tablecaption{Input Parameters for Figure 10 \label{table2}}
\tablewidth{0pt}
\tablehead{
\colhead{ion} &
\colhead{$\lambda$(ref)} &
\colhead{$\tau$(ref)} & 
\colhead{v$_{min}$} &
\colhead{T$_{exc}$}
} 
\startdata
Si II  &$\lambda$6347 &3.5  &      & 8000 \\
S II   &$\lambda$5454 &1.6  &      &10000 \\
Ca II  &$\lambda$3934 &6.0  &      & 5000 \\
Si III &$\lambda$4553 &3.0  &      &14000 \\
Fe III &$\lambda$5156 &1.3  &      &16000 \\
Fe II  &$\lambda$5018 &0.4  &20000 & 7000 \\
C II   &$\lambda$6578 &0.9 &14000 &12000 \\
\enddata
\end{deluxetable}
\clearpage

\begin{deluxetable}{crcrr}
\footnotesize
\tablecaption{Input Parameters for Figure 18 \label{table3}}
\tablewidth{0pt}
\tablehead{
\colhead{ion} &
\colhead{$\lambda$(ref)} &
\colhead{$\tau$(ref)} & 
\colhead{v$_{min}$} &
\colhead{T$_{exc}$}
} 
\startdata
Si II  &$\lambda$6347 & 4.0  &12000 & 8000 \\
S II   &$\lambda$5454 & 1.5  &12000 &10000 \\
Ca II  &$\lambda$3934 &  30  &      & 5000 \\
Si III &$\lambda$4553 & 2.2  &      &14000 \\
Fe III &$\lambda$5156 & 1.7  &      &14000 \\
Fe II  &$\lambda$5018 & 0.4  &18000 & 7000 \\
C II   &$\lambda$6578 & 0.5 &      &12000 \\
\enddata
\end{deluxetable}
\clearpage

\begin{deluxetable}{crcrr}
\footnotesize
\tablecaption{Input Parameters for Figure 19 \label{table4}}
\tablewidth{0pt}
\tablehead{
\colhead{ion} &
\colhead{$\lambda$(ref)} &
\colhead{$\tau$(ref)} & 
\colhead{v$_{min}$} &
\colhead{T$_{exc}$}
} 
\startdata
Si II  &$\lambda$6347 &12   &      & 8000 \\
S II   &$\lambda$5454 &1.8  &      &10000 \\
Ca II  &$\lambda$3934 &40   &      & 5000 \\
Fe III &$\lambda$5156 &1.2  &      &14000 \\
Fe II  &$\lambda$5018 &2.0  &      & 7000 \\
Fe II  &$\lambda$5018 &2.0  &18000 & 7000 \\
C II   &$\lambda$6578 &0.9  &      &12000 \\
Mg II  &$\lambda$4481 &1.0  &      &12000 \\
Na I   &$\lambda$5890 &0.5  &      &12000 \\
Co II  &$\lambda$4161 &1.0  &14000 &12000 \\
Ti II  &$\lambda$4550 &0.2  &14000 & 6000 \\
\enddata
\end{deluxetable}

\eject
\end{document}